\documentclass[nojss]{jss}
\usepackage{subfig}

\author{Paul Fearnhead\\Lancaster Universtity \And 
        Daniel Grose\\Lancaster University}
\title{\pkg{cpop}: Detecting changes in piecewise-linear signals}

\Plainauthor{Paul Fearnhead, Daniel Grose} %% comma-separated
\Plaintitle{\pkg{cpop}: Detecting changes in piecewise-linear signals} %% without formatting
\Shorttitle{\pkg{cpop}: Changes in slope} %% a short title (if necessary)

\Abstract{
Changepoint detection is an important problem with applications across many application domains. There are many different types of changes that one may wish to detect, and a wide-range of algorithms and software for detecting them. However there are relatively few approaches for detecting changes-in-slope in the mean of a signal plus noise model. We describe the \proglang{R} package, \pkg{cpop}, available on the Comprehensive R Archive Network (CRAN). This package implements CPOP, a dynamic programming algorithm, to find the optimal set of changes that minimises an $L_0$ penalised cost, with the cost being a weighted residual sum of squares. The package has extended the CPOP algorithm so it can analyse data that is unevenly spaced, allow for heterogeneous noise variance, and allows for a grid of potential change locations to be different from the locations of the data points. There is also an implementation that uses the CROPS algorithm to detect all segmentations that are optimal as you vary the $L_0$ penalty for adding a change across a continuous range of values.
}
\Plainkeywords{changepoints, change-in-slope, dynamic programming, piecewise linear models, structural breaks} %% without formatting
\Address{
  Paul Fearnhead\\
  Department of Mathematics and Statistics\\
  Lancaster University\\
  Lancaster,  \\
  LA1 4YF, UK\\
  E-mail: \email{p.fearnhead@lancaster.ac.uk}\\ ~\\
  Daniel Grose\\
  Department of Mathematics and Statistics\\
  Lancaster University\\
  Lancaster,  \\
  LA1 4YF, UK\\
  E-mail: \email{dan.grose@lancaster.ac.uk}\\
  
}
\begin{document}

%% include your article here, just as usual
%% Note that you should use the \pkg{}, \proglang{} and \code{} commands.

\section[Introduction]{Introduction}
%% Note: If there is markup in \(sub)section, then it has to be escape as above.

The detection of change in sequences of data is important across many applications, for example changes in volatility in finance \citep{andreou2002detecting}, changes in genomic data that represent copy number variation \citep{niu2012screening}, changes in calcium imaging data that correspond to neurons firing \citep{jewell2020fast} or changes in climate data \citep{reeves2007review}, amongst many others.  Depending on the application, interest can be in detecting changes in different features of the data, and there has been a corresponding wide-range of methods that have been developed. See \cite{aminikhanghahi2017survey}, \cite{truong2020selective}, \cite{fearnhead2020relating} and \cite{shi2022comparison} for recent reviews of changepoint methods and their applications.

\begin{figure}
\centering
\includegraphics[scale=1.4]{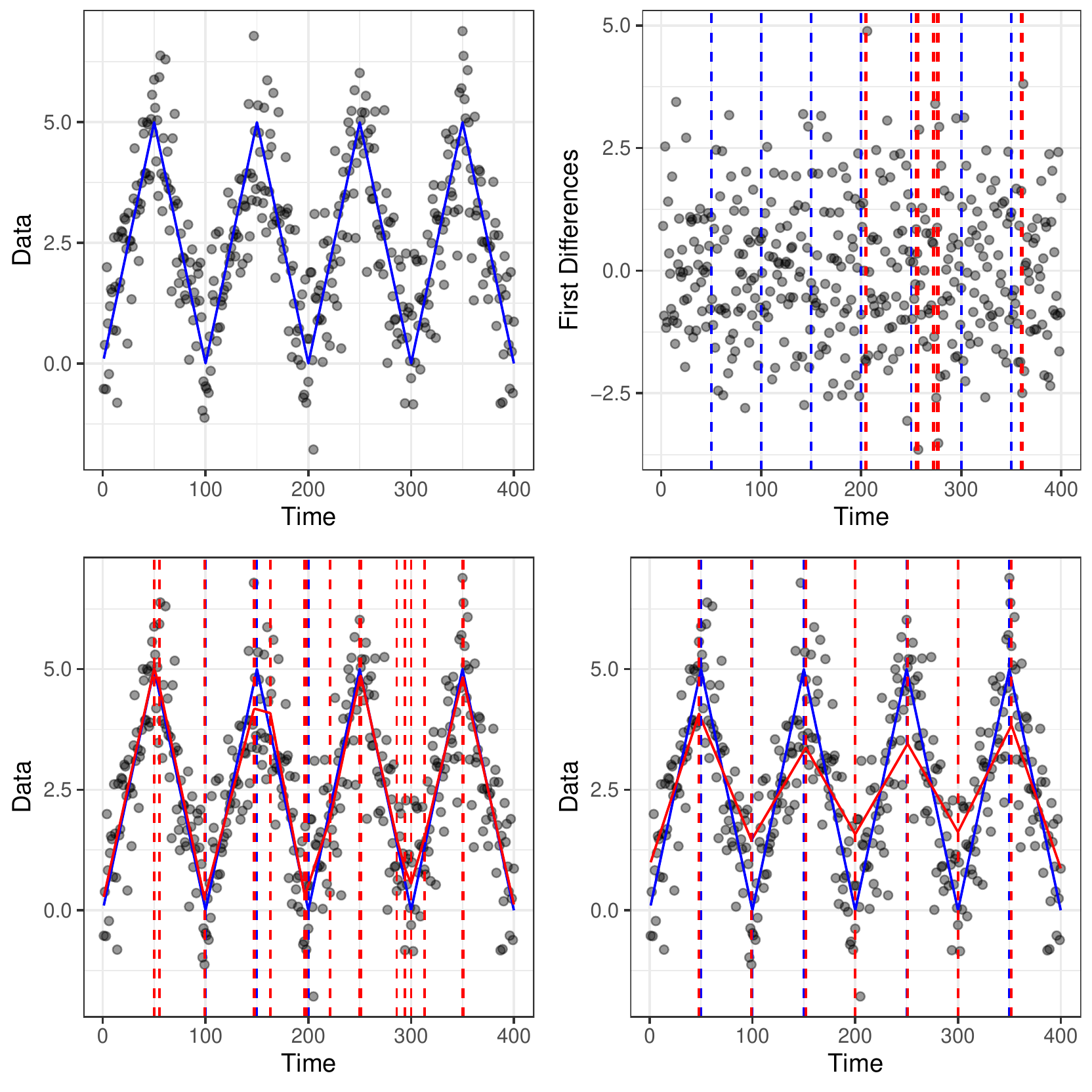}
\caption{ Example data simulated from a change-in-slope model (top left), and results from applying a change-in-mean algorithm to the first differences (top right) or from using trend-filtering (bottom row). In each case the true mean function (solid line) and change locations (vertical dashed lines) are shown in blue, and the estimates in red.  For trend filtering we chose the $L_1$ penalty value based on cross-validation (bottom left) or so that it obtained the correct number of changes (bottom right). In the former case we over-estimate the number of changes, while in the latter we obtain a poor estimate of the mean.
\label{fig:intro}}
\end{figure}

For some applications we have data on a piece-wise linear mean function, and we wish to detect the times at which the slope of the mean changes. This is the change-in-slope problem: see the top-left plot of Figure \ref{fig:intro} for example simulated data.  This is a particularly challenging problem for the following reasons. First, a simple approach to detecting changes in slope is to take first differences of the data, as this transforms a change-in-slope into a change-in-mean, and then apply one of the many methods for detecting changes in mean. However this removes much of the information in the data about the location of changes and such approach can perform poorly. This can be seen by comparing the raw data in the top-left plot of Figure \ref{fig:intro} with the first differenced data in the top-right plot Figure \ref{fig:intro}. By eye it is easy to see the rough location of the changes in slope in the former, but almost impossible to see any changes in mean in the latter. Running the PELT change-in-mean algorithm \cite{killick2012optimal} on the first differenced data leads to poor estimates of the location of any changes. Second, the most common approach to detecting multiple changepoints is to use binary segmentation \citep{scott1974cluster} or one of its variants \citep{fryzlewicz2014wild,kovacs2020seeded}. These repeatedly apply a test for a single change-in-slope. However \cite{baranowski2016narrowest} shows that such binary segmentation methods do not work for the change-in-slope problem as if you fit a single change-in-slope to data simulated with multiple changes, it will often detect the change at a location near the middle of a segment between changes. Third, dynamic programming algorithms that minimise an $L_0$ penalised cost, such as Optimal Partitioning \citep{jackson2005algorithm} or PELT \citep{killick2012optimal} cannot be applied to the change-in-slope problem due to dependencies in the model across changepoints from the continuity of the mean at each change. 

Despite these challenges, there are three methods developed specifically for detecting changes-in-slope: Trend-filtering \citep{kim2009ell_1,tibshirani2014adaptive} which minimises the residual sum of squares of fit to the data plus an $L_1$ penalty on the changes-in-slope; NOT \citep{baranowski2016narrowest} that repeatedly performs a test for a single change-in-slope on subsets of the data and combines the results using the narrowest-over-threshold procedure; and CPOP \citep{fearnhead2019detecting} which uses a novel variant of dynamic programming to minimise the residual sum of squares plus an $L_0$ penalty, i.e. a constant penalty for adding each change. The difference between the $L_1$ penalty of trend-filtering and the $L_0$ penalty of CPOP is important in practice: as the former allows one to fit a single change in slope with multiple changes of the same sign. This can lead to either over-fitting the number of changes, or, if a large enough penalty is used to detect the changes accurately, over-smoothing the mean function: see the bottom row of plots in Figure \ref{fig:intro} for an example. The main difference between CPOP and NOT is that the former fits all changes simultaneously. See \cite{fearnhead2019detecting} for an empirical comparison of the three methods.

The purpose of this paper is to describe the \pkg{cpop} package, which is written in \proglang{R}, and implements the CPOP algorithm. The latest version of the package was developed in response to an applied challenge, see Section \ref{sec:app}, where the data was unevenly spaced and the noise was not homoscedastic, aspects that previous implementations of change-in-slope algorithms could not handle. How the CPOP algorithm is extended to deal with these features is described in Section \ref{sec:background}, together with allowing the locations of the changes in slope to not coincide with the observations. This latter aspect can be helpful in reducing the computational cost of the CPOP algorithm for high frequency data by, e.g., searching for segmentations that only allow changes at a smaller grid of possible locations. Section \ref{sec:cpop} describes the basic functionality of the package, with the extensions to allow for unevenly spaced, heteroscedastic data described, and to specify the grid of potential change locations, in Section \ref{sec:extensions}. This latter section also shows how to impose a minimum segment length and how to implement CPOP within the CROPS algorithm \citep{haynes2017computationally} to obtain all segmentations as we vary the value of $L_0$ penalty. An application of CPOP to analyse decay of spectra from ocean models is shown in Section \ref{sec:app}.

\subsection{R Packages for Changepoint Detection}

There are both many different types of change that one may wish to detect, and many different approaches to detecting multiple changes. Consequently there are a wide range of change algorithms with associated packages in \proglang{R}. For example the \pkg{changepoint} package \citep{killick2014changepoint} implements dynamic programming algorithms, such as PELT, for detecting changes in mean, variance or mean and variance. Other dynamic programming algorithms include \pkg{fpop} and \pkg{gfpop} \citep{runge2020gfpop} that implements the functional optimal partitioning algorithm \citep{maidstone2017optimal} for detecting changes in mean, with the latter package allowing for flexibility as to how the mean changes (such as monotonically increasing) and for different loss functions for measuring fit to data. The \pkg{breakfast} package implements a range of methods based on recursively applying a test to detect a single change, for example wild binary segmentation \citep{fryzlewicz2014wild} and IsolateDetect \citep{anastasiou2022detecting}; whilst \pkg{mosum} \citep{meier2021mosum} implements the MOSUM procedure. Packages \pkg{stepR} and \pkg{FDRseg} implement the multiscale approaches of \cite{frick2014multiscale}, \cite{pein2017heterogeneous} and \cite{li2016fdr}.

Separately there are packages that perform non-parametric change detection, for example \pkg{ecp} \citep{james2015ecp} implements the method of \cite{matteson2014nonparametric}, while \pkg{changepoint.np}  implements the method of \cite{haynes2017NP}. There are also methods for analysing multiple dimensional data streams, such as \pkg{InspectChangepoint} \citep{wang2018high}, %\pkg{changepoint.mv} \citep{}
and \pkg{changepoint.geo} \citep{grundy2020high}; Bayesian methods, such as \pkg{bcp} \citep{erdman2008bcp}; and methods that implement online procedures such as \pkg{CPM} \citep{ross2015parametric} and \pkg{FoCUS} \citep{romano2021fast}.

However, as mentioned above, there are more limited methods for specifically detecting changes-in-slope. The trend filtering algorithm can be implemented using the \code{trendfilter} function from the \pkg{genlasso} \citep{genlasso} package, and the NOT algorithm can be implemented using the \pkg{not} package or is available within \pkg{breakfast}. However current implementations of these do not allow for unevenly spaced, heterogeneous observations or minimum segment lengths. (Though there is flexibility within the \pkg{genlasso} package for implementing general lasso algorithms, and these can be constructed to fit a trend-filtering model to unevenly spaced data).

\section[Background]{Detecting changes in slope} \label{sec:background}

%We are interested in the change-in-slope problem.
Assume we have data points $(y_1,x_1),\ldots,(y_n,x_n)$, ordered so that $x_1<x_2<\cdots<x_n$. In many applications $x_i$ will be a time-stamp of when response $y_i$ is obtained, whilst in, say, genomic applications, $x_i$ may correspond to a location along the genome at which observation $y_i$ is taken. We wish to model the response, $y$, as a signal plus noise where the signal is modelled as a continuous piecewise linear function of $x$. That is
\begin{equation} \label{eq:cinslope}
y_i=f(x_i)+\epsilon_i,
\end{equation}
where $f(x)$ is a continuous piecewise linear function, and $\epsilon_i$ is noise. If the function $f(x)$ has $K$ changes in slope in the open interval $(x_1,x_n)$, and these occur at $x$-values $\tau_1,\ldots,\tau_K$, and we define $\tau_0$ and $\tau_{K+1}$ to be arbitrary values such that $\tau_0\leq x_1$ and $\tau_{K+1}\geq n$ then we can uniquely define $f(x)$ on $[x_1,x_n]$ by specifying the values $f(\tau_i)$ for $i=0,\ldots,K+1$. The function $f(x)$ can then be obtained via straight-line interpolation between the points $(\tau_i,f(\tau_i))$.

Our interest is in estimating the number of changes in slope, $K$, their locations, $\tau_1,\ldots,\tau_K$, and the underlying signal. The latter is equivalent to estimating $f(\tau_i)$ for $i=0,\ldots,K+1$. To simplify notation we will denote these values by $\alpha_0,\ldots,\alpha_{K+1}$, i.e.~ $\alpha_i=f(\tau_i)$ for $i=0,\ldots,K+1$. Also, for this and other quantities we will use the shorthand $\alpha_{i:j}$ for integers $i\leq j$ to be the ordered set of values, $\alpha_i,\ldots,\alpha_j$.

\subsection{An $L_0$-Penalised Criteria}

To estimate the number and locations of the changes-in-slope, and the underlying signal, we will first introduce a grid of $x$-values, $g_{1:N}$ with these ordered so that $g_i<g_j$ if and only if $i<j$. Our estimate for $f(x)$ will be restricted to piecewise-linear functions whose slope is only allowed to change at these grid-points. We will define our estimator of $f(x)$ as the function that minimises a penalised cost that is a sum of the fit of the function to data, measured in terms of a weighted residual sum of squares, plus a penalty for each change-in-slope. That is we solve the following minimisation problem
\begin{equation} \label{eq:penalised_cost}
\min_{K,\tau_{1:K}\in g_{1:N}, \alpha_{0:K+1} } \left\{
\sum_{i=1}^n \frac{1}{\sigma^2_i} \left(y_i -  \alpha_{j(i)}-(\alpha_{j(i)+1}-  \alpha_{j(i)})\frac{x_i-\tau_{j(i)}}{\tau_{j(i)+1}-\tau_{j(i)}}   \right)^2
+K\beta
\right\},
\end{equation}
where $\beta>0$ is a user chosen penalty for adding a changepoint, $j(i)$ is such that $\tau_{j(i)}\leq x_i < \tau_{j(i)+1}$, and $\sigma^2_{1:n}$ are user specified constants that are estimates of the variances of the noise $\epsilon_{1:n}$. The cost that we are minimising consists of two terms. The first is the measure of fit to the data, and is a residual sum of squares, but with the residuals weighted by the inverse of the variance of the noise for that observation. The expression in this term that depends on $\alpha_{0:K+1}$ and $\tau_{0:K+1}$ is just an expression for $f(x_i)$ given that $f(x)$ is defined as the linear interpolation between the points $(\tau_i,\alpha_i)$ for $i=0,\ldots,K+1$. The second term is the penalty for the number of changes-in-slope, with a penalty of $\beta$ for each change.

This approach for estimating changes-in-slope was first proposed in \cite{fearnhead2019detecting}, but they assumed that the locations of the data points were evenly spaced, so $x_i=i$, the grid-points were equal to the locations of the data points, so $N=n$ and $g_{1:N}=x_{1:n}$, and that the noise was homogeneous so $\sigma^2_i=\sigma^2$, for some constant $\sigma^2$, for all $i$. 

Before we describe how to extend the approach in \cite{fearnhead2019detecting} to this more general estimator, we first give some comments on this and related approaches to estimating changes-in-slope. This approach is a common for estimating changes \citep{jackson2005algorithm,killick2012optimal} and the cost is often termed an $L_0$ penalised cost. This is to contrast it with $L_1$ penalised costs, such as implemented in trend-filtering \citep{kim2009ell_1,tibshirani2014adaptive} which are of similar form, except that the cost for adding a change-in-slope is linear in the size of the change-in-slope. The advantage of using an $L_1$ penalised cost is that solving the resulting minimisation problem is simpler -- however such an approach tends to over-estimate the number of changes, as shown in the introduction. An alternative approach to estimating changes-in-slope is to perform tests for a change-in-slope on data from randomly chosen intervals of $x$-values and to combine the results of these tests using the narrowest-over-threshold procedure of \cite{baranowski2016narrowest}. The current formulation and implementation of these alternative methods also make the simplifying assumptions of evenly spaced locations for the data, change-in-slope only at data point locations, and that the noise variance is constant. %For evenly spaced data points, for example when $x_i=i$, it may seem natural to take first differences of the data, and convert the change-in-slope problem into a change-in-mean problem. We can then apply one of the many methods for detecting changes in mean. This approach is not to be recommended, as these methods ignore the auto-correlation that is induced into the residuals of the differenced data, and it can be shown to have a much lower power than solving the change-in-slope problem directly. We explore this empirically in Section \ref{sec:comparison_evenly_spaced}.

The choice of $\beta$ in (\ref{eq:penalised_cost}) is important for accurate estimates, with lower values of $\beta$ leading to larger estimates of $K$, the number of changes. If the noise is approximately Gaussian and independent, and the estimate of the noise variance is good, then $\beta=2\log n$ is an appropriate choice  \citep{fearnhead2019detecting}. In general these assumptions will not hold, and often larger values of $\beta$ are required to e.g. compensate for positively auto-correlated noise. Where possible we recommend evaluating sets of estimated changepoints obtained for a range of $\beta$ values, and these can obtained in a computationally efficient manner using the CROPS algorithm of \cite{haynes2017computationally}.

\subsection{Dynamic Programming Recursion}

Solving (\ref{eq:penalised_cost}) is non-trivial as it involves minimising a non-convex function. Furthermore, standard dynamic programming algorithms for change points \citep{maidstone2017optimal}, e.g. those that recurse based on conditioning on the location of the most recent changepoint,  cannot be used because of the dependence across changepoints due to the continuity constraint. Thus we follow \cite{fearnhead2019detecting} and develop a dynamic programming recursion that conditions both on the location of the changepoints and the value of the mean at that changepoint.

Remember that we are allowing changepoints only at grid-points, $g_{1:N}$. We introduce a set of functions, each associated with a grid-point, $F_t(\alpha)$ for $t=1,\ldots,N$, defined as the minimum value of the penalised cost for data up to and including grid-point $g_t$ conditional on the signal at $g_t$ being $\alpha$, i.e. $f(g_t)=\alpha$. To define this formally, define a set of segment cost functions
\[
\mathcal{C}_{s,t}(\alpha',\alpha) = \sum_{i=n_s+1}^{n_t} \left(y_i -\alpha' - (\alpha-\alpha')\frac{x_i-g_s}{g_t-g_s} \right)^2,
\]
which is the cost of fitting a linear signal to data points between the $s$th and $t$th grid-points, i.e. $(x_i,y_i)$ with $g_s<x_i\leq g_t$, with the signal taking the value $\alpha'$ at $g_s$ and $\alpha$ at $g_t$. In the summation we use the notation $n_s$ to denote the index of the last data-point located at or before $g_s$. If $n_t=n_s$, that is there are no data-points between $g_s$ and $g_t$ then this cost is set to 0.

Using this definition, the $L_0$ penalised criteria (\ref{eq:penalised_cost}) can be written as
\begin{equation} \label{eq:penalisedcost2}
\min_{K,i_{1:K}\in {1:N}, \alpha_{0:K+1} }  \left\{
\sum_{k=0}^K \mathcal{C}_{i_k,i_{k+1}}(\alpha_k,\alpha_{k+1})
+K \beta
\right\}.
\end{equation}
Using this definition can define the function $F_l(\alpha)$ for $l=1,\ldots,N$ as
\[
F_l(\alpha)=
\min_{K,i_{1:K}\in {1:l-1}, \alpha_{0:K} } \left\{
\sum_{k=0}^{K-1} \mathcal{C}_{i_k,i_{k+1}}(\alpha_k,\alpha_{k+1})
+K\beta
+\mathcal{C}_{i_K,l}(\alpha_K,\alpha)
\right\},
\]
which is of the form of (\ref{eq:penalisedcost2}) but with $\tau_{K+1}=g_l$ and $\alpha_{K+1}=\alpha$, as for $F_l(\alpha)$ we are analysing only data up to $g_l$ and we are fixing $\alpha_{K+1}=f(g_l)=\alpha$.

Using the same argument as in \cite{fearnhead2019detecting} we can then derive a recursion for $F_l(\alpha)$. For $l=1,\ldots,N$,
\[
F_l(\alpha)= \min_{k \in 0:(l-1)} \left\{
\min_{\alpha'} \left[
F_{k}(\alpha')+\mathcal{C}_{k,l}(\alpha',\alpha) +\beta
\right]
\right\},
\]
with $F_0(\alpha)= -\beta$. The idea of this recursion is that we condition on the location of the most recent changepoint, $g_k$, and the value of the signal at that changepoint, $\alpha'$. Conditional on this information the optimal segmentation prior to the most recent changepoint is independent of the data since that changepoint, and the minimum of the penalised cost is $F_{k}(\alpha')+\mathcal{C}_{l-1,l}(\alpha',\alpha) +\beta$: the sum of the minimum cost for the data prior to $g_k$, plus the segment cost for the data from $g_k$ to $g_l$ plus the penalty for adding a changepoint. Finally we obtained $F_l(\alpha)$ by minimising over $k$ and $\alpha'$.

Solving this recursion is possible as the functions $F_l(\alpha)$ can be summarised as the pointwise minimum of a set of quadratics. For each $k$, we can solve the inner minimisation over $\alpha'$ analytically. Doing so for each $k$ will define $F_l(\alpha)$ as the pointwise minimum of a large set of quadratics, and we can use a line search to then prune quadratics that do not contribute to this minimum (which is important to reduce the computational cost of the algorithm). See \cite{fearnhead2019detecting} for full details. As noted there, it is possible to further reduce the computational cost of solving the recursion by using PELT pruning ideas from \cite{killick2012optimal}. This pruning enables us to reduce the search space of $k$ within the recursion. Finally, whilst we have described how to find the minimum value of the penalised cost, our main interest is in the locations of the changepoints and the value of the signal that gives that minimum cost. However extracting this information is trivial once we have solved the recursions -- again see \cite{fearnhead2019detecting} for details.

The novelty relative to \cite{fearnhead2019detecting} is that for this recursion we have decoupled the grid of potential changepoints from the locations of the datapoints. Furthermore, our setting allows for unevenly spaced data and for the noise variance to be heterogeneous. These impact that definition of $\mathcal{C}_{l-1,l}(\alpha',\alpha)$ and how we perform the inner minimisation over $\alpha'$. Full details are given in Appendix \ref{App:cost_update}.

One further extension of this approach to detecting changes is to allow for a minimum segment length. This can be done by optimising the penalised cost (\ref{eq:penalised_cost}) only over sets of changepoints that are consistent with the prescribed minimum segment length. To minimise the cost subject to such a constraint involves adapting the dynamic programming recursion so that we only search over values of $k$ for the location of the most recent changepoint that are consistent with the minimum segment length. However one drawback with imposing a minimum segment length for the change-in-slope problem is that it makes the PELT pruning ideas invalid. In our implementation of \code{cpop}, if a minimum segment length is specified we allow the algorithm to be run both with and without the PELT pruning. If run without PELT pruning, the algorithm will be slower but guaranteed to find the optimal segmentation under our condition. Running with PELT pruning is quicker but the algorithm may output a slightly sub-optimal segmentation. In practice we have observed that this happens rarely. %, and we investigate this empirically in Section \ref{}.

\section{The {cpop} Package} \label{sec:cpop}

The \pkg{cpop} package has functions to simulate data from a change-in-slope model, implement the CPOP algorithm to estimate the location of the changes, and various functions for summarising and plotting the estimates of the change locations and the mean function. 

\subsection{Generating Simulated Data}
The \code{simchangeslope} function allows for simulating data from a change-in-slope model (\ref{eq:cinslope}): %. It takes as input the $x$-values of the data points, the locations of the changes, how much the slop of the mean changes at each changepoint, and the standard deviation of the noise:
\begin{CodeInput}
simchangeslope(x, changepoints, change.slope, sd = 1)
\end{CodeInput}
It takes the following arguments :
\begin{itemize}
\item \code{x} -  A numeric vector containing the locations of the data.
\item \code{changepoints} - A numeric vector of changepoint locations.
\item \code{change.slope} -  A numeric vector indicating the change in slope at each changepoint. The initial slope is assumed to be 0.
\item \code{sd} - The residual standard deviation. Can be a single numerical value or a vector of values for the case of varying residual standard deviation. Default value is 1.
\end{itemize}
It returns a vector $y$ of simulated values which correspond to the locations $x$. 
The mean function of the data goes through the origin -- but to add an intercept we just add a constant to all output values. It is possible to get the value of the mean function at the $x$-values of the data by setting \code{sd=0}.

\begin{figure}
\centering
\includegraphics[width=0.6\linewidth]{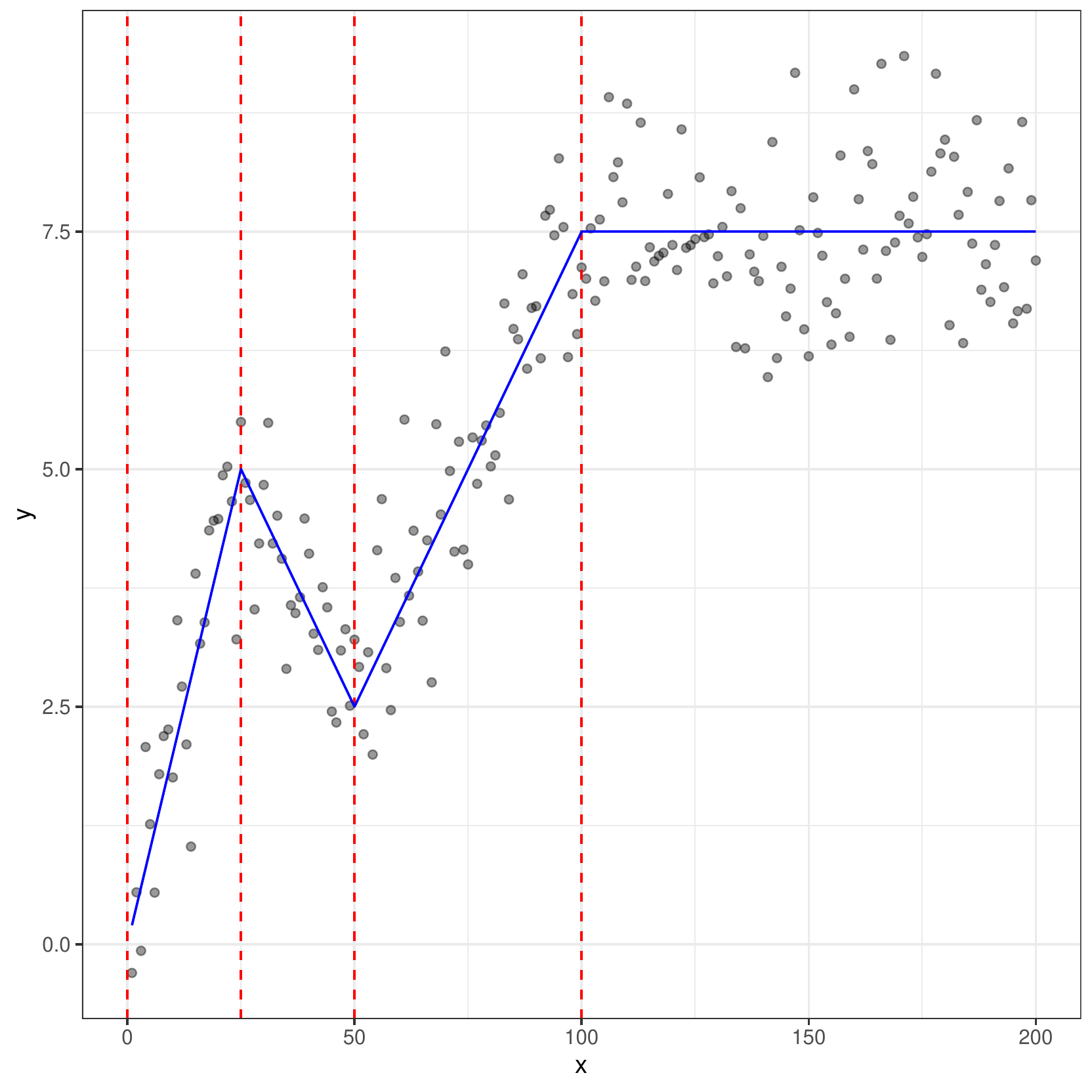}
\caption{Simulated data (black dots) with true mean (blue dashed line) and changepoints (vertical red dashed lines).}
\label{fig:simulate-example}
\end{figure}

The following code demonstrates the \code{simchangeslope} function and displays the data along with the (true) line segments and the locations of the changes in slope (see Figure \ref{fig:simulate-example}).
\begin{CodeChunk}
\begin{CodeInput}
R> library("cpop")
R> library("ggplot2")
R> changepoints <- c(0, 25, 50, 100)
R> change.slope <- c(0.2, -0.3, 0.2, -0.1)
R> x <- 1:200
R> sd <- 0.8
R> y <- simchangeslope(x, changepoints, change.slope, sd)
R> df <- data.frame("x" = x, "y" = y)
R> p <- ggplot(data = df, aes(x = x, y = y))
R> p <- p + geom_point(alpha = 0.4)
R> p <- p + geom_vline(xintercept = changepoints,
+                      color = "red",
+                      linetype = "dashed")
R> mu <- simchangeslope(x, changepoints, change.slope, sd = 0)
R> p <- p + geom_line(aes(y = mu), color = "blue")
R> p <- p + theme_bw()
R> print(p)
\end{CodeInput}
\end{CodeChunk}
\subsection{Determining Changes in Slope}
\begin{figure}
\centering
\includegraphics[width=0.6\linewidth]{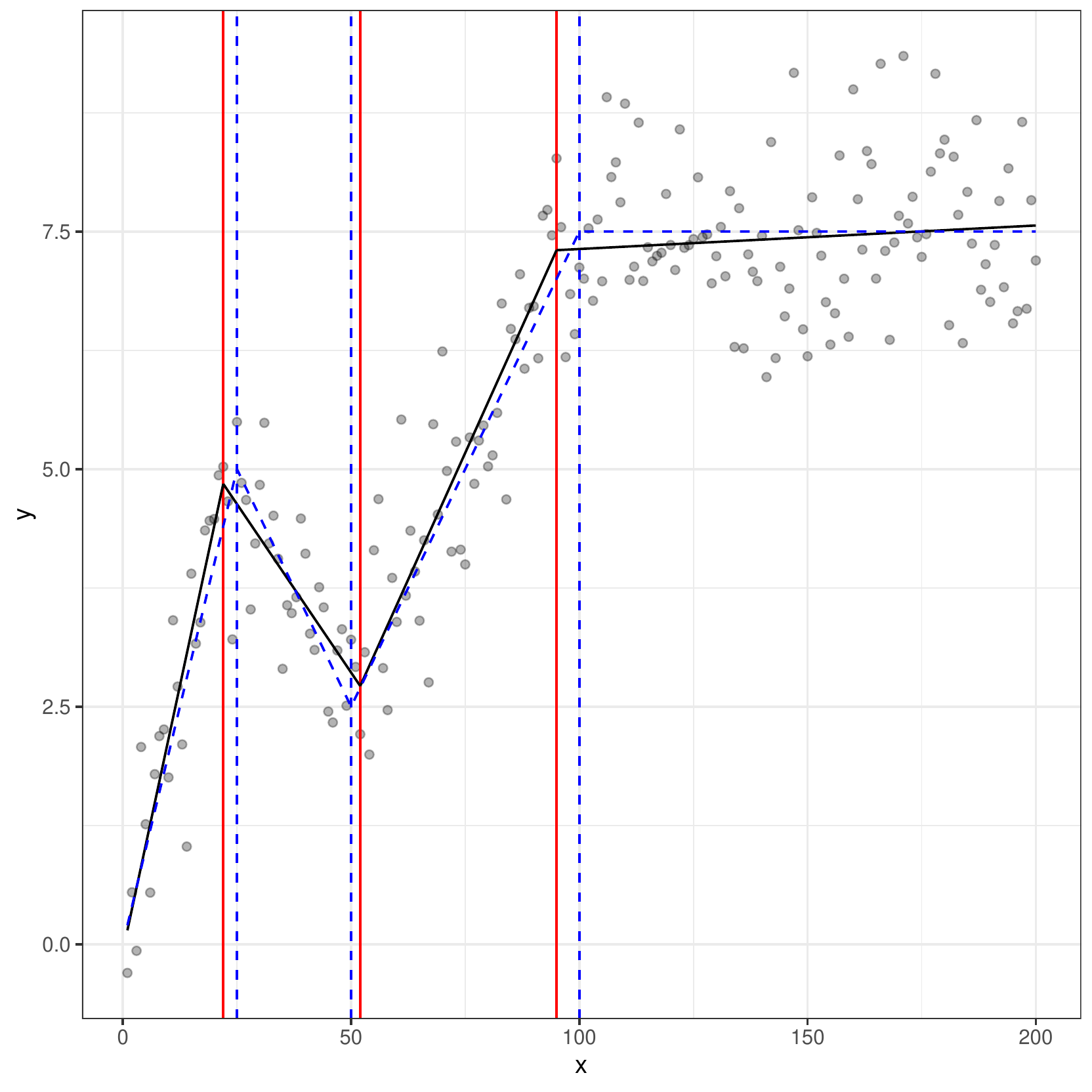}
\caption{Example output of \code{cpop} for simulated data from Figure \ref{fig:simulate-example}. The true mean and changepoints are given in blue dashed lines, together with estimated mean (black full line) and changepoints (red full lines).   }
\label{fig:cpop-example}
\end{figure}

The function \code{cpop} is used to determine the locations of changes in slope.
\begin{CodeInput}
cpop(y, x = 1:length(y) - 1, grid = x, beta = 2 * log(length(y)), 
    sd = 1, minseglen = 0, prune.approx = FALSE)
\end{CodeInput}
It takes the following arguments :
\begin{itemize}
 \item  \code{y} - A vector of length $n$ containing the data.
\item \code{x} -  A vector of length $n$ containing the times/locations of data points. Default value is NULL, in which case the locations are set to be $0,1,\ldots,n-1$, corresponding to evenly spaced data.
\item \code{grid} -  An ordered vector of possible locations for the change points. If this is NULL, then this is set to $x$, the vector of times/locations of the data points.
\item \code{beta} -  A positive real value for the penalty, $\beta$ in (\ref{eq:penalisedcost2}), incurred for adding a changepoint. The larger the penalty, the fewer changepoints will be detected.
\item \code{sd} - Estimate of residual standard deviation. Can be a single numerical value if it is the same for all data points, or a vector of $n$ values for the case of varying standard deviation. 
\item \code{minseglen} -  The minimum allowable segment length, i.e. distance between successive changepoints. The default is that no minimum segment length is imposed.
\item \code{prune.approx} -  Only relevant if a minimum segment length is set. If True, \code{cpop} will use an approximate pruning algorithm that will speed up computation but may occasionally lead to a sub-optimal solution in terms of the estimated changepoint locations. If the minimum segment length is 0, then an exact pruning algorithm is possible and is used.
\end{itemize}
The \code{cpop} function returns an S4 object for which a number of generic methods, including \code{plot} and \code{summary}, are provided. 
%
%
%\subsubsection{simulated data}
The following demonstrates how the \code{cpop} function can be used to determine changes in slope, by analysing the data we simulated and plotted in Figure \ref{fig:simulate-example}. It uses the default penalty, $\beta=2\log n$, and we assume that the true noise standard deviation, 0.8, is known. The \code{summary} function is used to provide an overview of the analysis parameters along with estimated changepoint locations, corresponding fitted line segments, and the (weighted) residual sum of squares (RSS) for each segment. 
\begin{CodeChunk}
\begin{CodeInput}
R> res <- cpop(y, x, sd = 0.8)
R> summary(res)
\end{CodeInput}
\begin{CodeOutput}
cpop analysis with n = 200 and penalty (beta)  = 10.59663

3  changepoints detected at x = 
 22 52 95
fitted values : 
  x0       y0  x1       y1     gradient   intercept      RSS
1  1 0.147335  22 4.844725  0.223685242 -0.07635023 10.07761
2 22 4.844725  52 2.717661 -0.070902123  6.40457180 10.38813
3 52 2.717661  95 7.303644  0.106650750 -2.82817758 25.09463
4 95 7.303644 200 7.563413  0.002473995  7.06861408 61.78303

overall RSS = 107.3434
cost = 199.514
\end{CodeOutput}
\end{CodeChunk}
The predicted change in slope (changepoint) locations and corresponding line segments can be displayed using \code{plot}. The \code{plot} function returns a \pkg{ggplot2} object which can be augmented to include additional features such as the true change in slope locations and line segments (see Figure \ref{fig:cpop-example}).
\begin{CodeChunk}
\begin{CodeInput}
R> p <- plot(res)
R> p <- p + geom_vline(xintercept = changepoints[-1], color = "blue",
+                   linetype = "dashed")
R> p <- p + geom_line(aes(y = mu), color = "blue", linetype = "dashed")
R> print(p)
\end{CodeInput}
\end{CodeChunk}
The last two lines of code add the true changepoint locations and the true mean function to the plot. For plotting the changepoint location we omit the first element of \code{changepoints}, which was 0, as that was included just to set the initial slope of the mean and does not correspond to a change-in-slope.

\subsection{Other Functions}
In addition to \code{plot} and \code{summary}, the \pkg{cpop} package provides functions to evaluate the fitted mean function at specified $x$-values, and to calculate the residuals of the fitted mean. The primary argument of these functions is \code{object} - An instance of a cpop S4 class as produced by the function \code{cpop}.

The function \code{changepoints(object)}
creates a data frame containing the locations of the changepoints in terms of the their $x$-values.
\begin{CodeChunk}
\begin{CodeInput}
R> changepoints(res)
\end{CodeInput}
\begin{CodeOutput}
  location
1       22
2       52
3       95
\end{CodeOutput}
\end{CodeChunk}
%
%
%The function \code{cost(object)} calculates the penalised cost of a model fitted by \code{cpop} using the residual sum of squares and the penalty values.
%\begin{CodeChunk}
%\begin{CodeInput}
%R> cost(res)
%\end{CodeInput}
%\begin{CodeOutput}
%[1] 47.39048
%\end{CodeOutput}
%\end{CodeChunk}
%
%
The function \code{estimate(object, x = object@x, ...)} with argument,
 \code{x}, that specifies the $x$-values at which the fit is to be estimated, creates a data frame with two columns containing the locations \code{x} and the corresponding estimates $\hat{y}$. The default value for \code{x} is the vector of $x$ locations at which the cpop object was defined.
  % \item \code{...} - For compatability with existing S4 generic methods. \end{itemize}%
%\subsubsection{example}
\begin{CodeChunk}
\begin{CodeInput}
R> estimate(res, x = c(0.1,2.7,51.6))
\end{CodeInput}
\begin{CodeOutput}
   x     y_hat
1  0.1 0.1473350
2  2.7 0.7289166
3 51.6 2.9473390
\end{CodeOutput}
\end{CodeChunk}
The function \code{fitted(object)} creates a data frame containing the endpoint coordinates for each line segment fitted between the detected
changepoints. The data frame also contains the gradient and intercept values for each segment and the corresponding residual sum of squares (RSS).
%\subsubsection{example}
\begin{CodeChunk}
\begin{CodeInput}
R> fitted(res)
\end{CodeInput}
\begin{CodeOutput}
  x0       y0  x1       y1     gradient   intercept      RSS
1  1 0.147335  22 4.844725  0.223685242 -0.07635023 10.07761
2 22 4.844725  52 2.717661 -0.070902123  6.40457180 10.38813
3 52 2.717661  95 7.303644  0.106650750 -2.82817758 25.09463
4 95 7.303644 200 7.563413  0.002473995  7.06861408 61.78303
\end{CodeOutput}
\end{CodeChunk}
Finally, the function \code{residuals(object)}  creates a single column matrix containing the residuals.
%\subsubsection{example}
\begin{CodeChunk}
\begin{CodeInput}
R> head(residuals(res))
\end{CodeInput}
\begin{CodeOutput}
           [,1]
[1,] -0.4484981
[2,]  0.1758944
[3,] -0.6632084
[4,]  1.2578339
[5,]  0.2215302
[6,] -0.7221359
\end{CodeOutput}
\end{CodeChunk}

\section{Extensions of {cpop}} \label{sec:extensions}

\subsection{Irregularly Sampled Data}

The \pkg{cpop} package allows for irregularly spaced data, both when simulating data and when running the CPOP algorithm. The only change to the previous code that we need to make is to change the definition of \code{x} that is input to \code{simchangeslope}.

\begin{CodeChunk}
\begin{CodeInput}
R> x <- (1:200)^(2)/(200)
R> changepoints <- c(0, 25, 50, 100)
R> change.slope <- c(0.2, -0.3, 0.2, -0.1)
R> sd <- 0.8
R> y <- simchangeslope(x, changepoints, change.slope, sd)
\end{CodeInput}
\end{CodeChunk}

\begin{figure}
\centering
\includegraphics[width=0.6\linewidth]{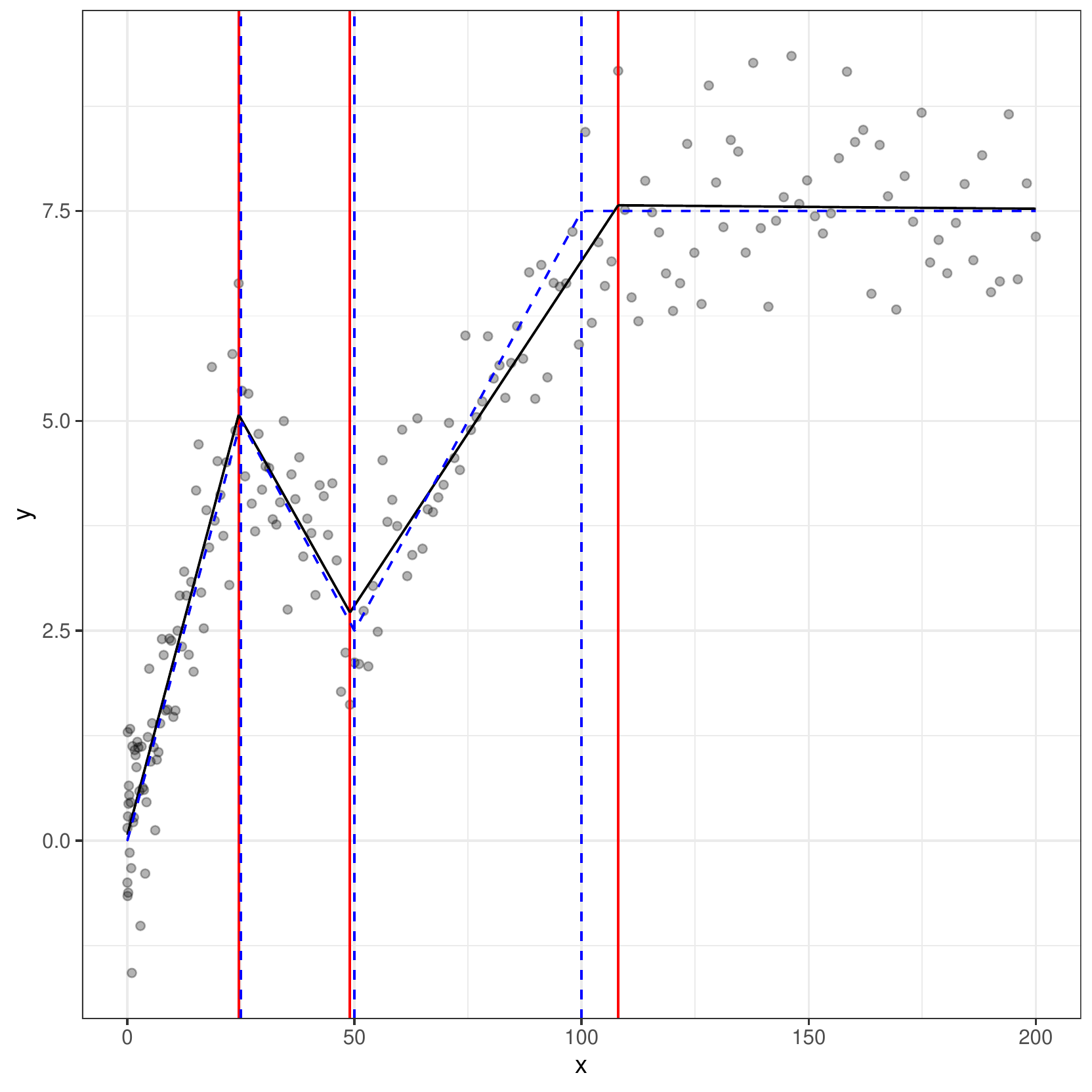}
\caption{Example output of \code{cpop} for unevenly spaced simulated data. The true mean and changepoints are given in blue dashed lines, together with estimated mean (black full line) and changepoints (red full lines). }
\label{fig:cpop-example-uneven}
\end{figure}

To analyse the data we use \code{cpop} as before. (The only difference is that for evenly spaced data one can omit the \code{x} argument -- but it must be included for unevenly spaced data.)
\begin{CodeChunk}
\begin{CodeInput}
R> res <- cpop(y, x, sd = 0.8)
\end{CodeInput}
\end{CodeChunk}
Figure \ref{fig:cpop-example-uneven} shows a plot of the simulated data and estimated changepoints and mean function.

\subsection{Heterogeneous Data}

To simulate heterogeneous data we just input a vector of the standard deviation of each data point. For example, we can produce a version of the simulation from Section \ref{sec:cpop} but with the noise standard deviation increasing with $x$.

\begin{CodeChunk}
\begin{CodeInput}
R> x <- 1:200
R> sd <- x / 100
R> y <- simchangeslope(x, changepoints, change.slope, sd))
\end{CodeInput}
\end{CodeChunk}
Here the values of \code{changepoints} and \code{change.slope} are as before.

\begin{figure}
\centering
{\includegraphics[scale=.5,width=14cm,trim={0 4cm 0 4cm},clip]{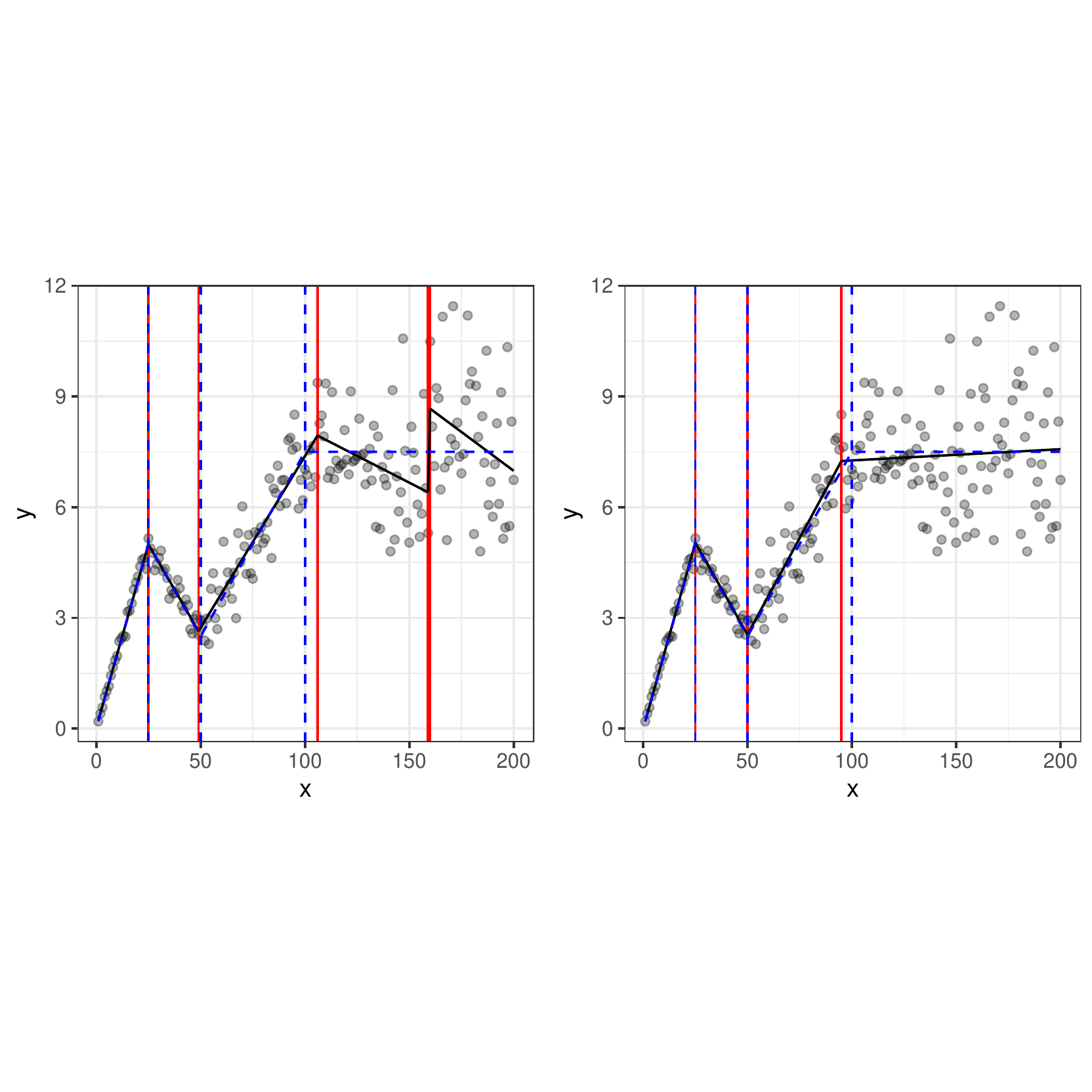}}
\caption{Example output of \code{cpop} for heterogenous noise: assuming a constant noise variance (left) and the true noise variance (right). The true mean and changepoints are given in blue dashed lines, together with estimated mean (black full line) and changepoints (red full lines). }
\label{fig:cpop-example-heterogeneous}
\end{figure}

It is interesting to compare two estimates of the changepoints, one where we assume a fixed noise standard deviation, and one where we assume the true noise standard deviation. For the former it is natural to set this value so the the average variance of the noise is correct. 
\begin{CodeChunk}
\begin{CodeInput}
R> res <- cpop(y, x, sd = sqrt(mean(sd^2)))
R> res.true <- cpop(y, x, sd = sd)
\end{CodeInput}
\end{CodeChunk}
Here \code{res} contains the results where we assume a fixed noise standard deviation, and \code{res.true} where we use the true values. Figure \ref{fig:cpop-example-heterogeneous} shows the results -- and we can see that wrongly assuming homogeneous noise leads to detecting two false positive changepoints in regions where the noise variance is above what was assumed. 

%\begin{figure}
%\begin{minipage}{.5\linewidth}
%\centering
%{\includegraphics[scale=.5]{figures/cpop-example-uneven2.pdf}}
%{\includegraphics[scale=.5]{figures/cpop_example_uneven_1_ggplot.pdf}}
%\end{minipage}%
%\begin{minipage}{.5\linewidth}
%\centering
%\includegraphics{figures/cpop-example-uneven.pdf}
%{\includegraphics[scale=.5]{figures/cpop_example_uneven_2_ggplot.pdf}}
%\end{minipage}%
%\caption{Example output of \code{cpop} for heterogenous noise: assuming a constant noise variance (left) and the true noise variance (right). The true mean and changepoints are given in blue dashed lines, together with estimated mean (black full line) and changepoints (red full lines). }
%\label{fig:cpop-example-heterogeneous}
%\end{figure}

One practical issue is how can we estimate the noise variance in the heterogeneous case? In some situations there may be covariate information that tells the relative variance of the noise for different data points (for example due to some data being averages of multiple measurements). Alternatively if we know how the noise variance depends on $x$ we can estimate this by (i) running CPOP assuming a constant variance; (ii) calculating the residuals of the fitted model; (iii) estimating how the noise variance varies with $x$ by fitting an appropriate model to the residuals. An example of this scheme will be seen in Section \ref{sec:app}.

\subsection{Choice of Grid}

\begin{figure}
\centering
\includegraphics[width=0.6\linewidth]{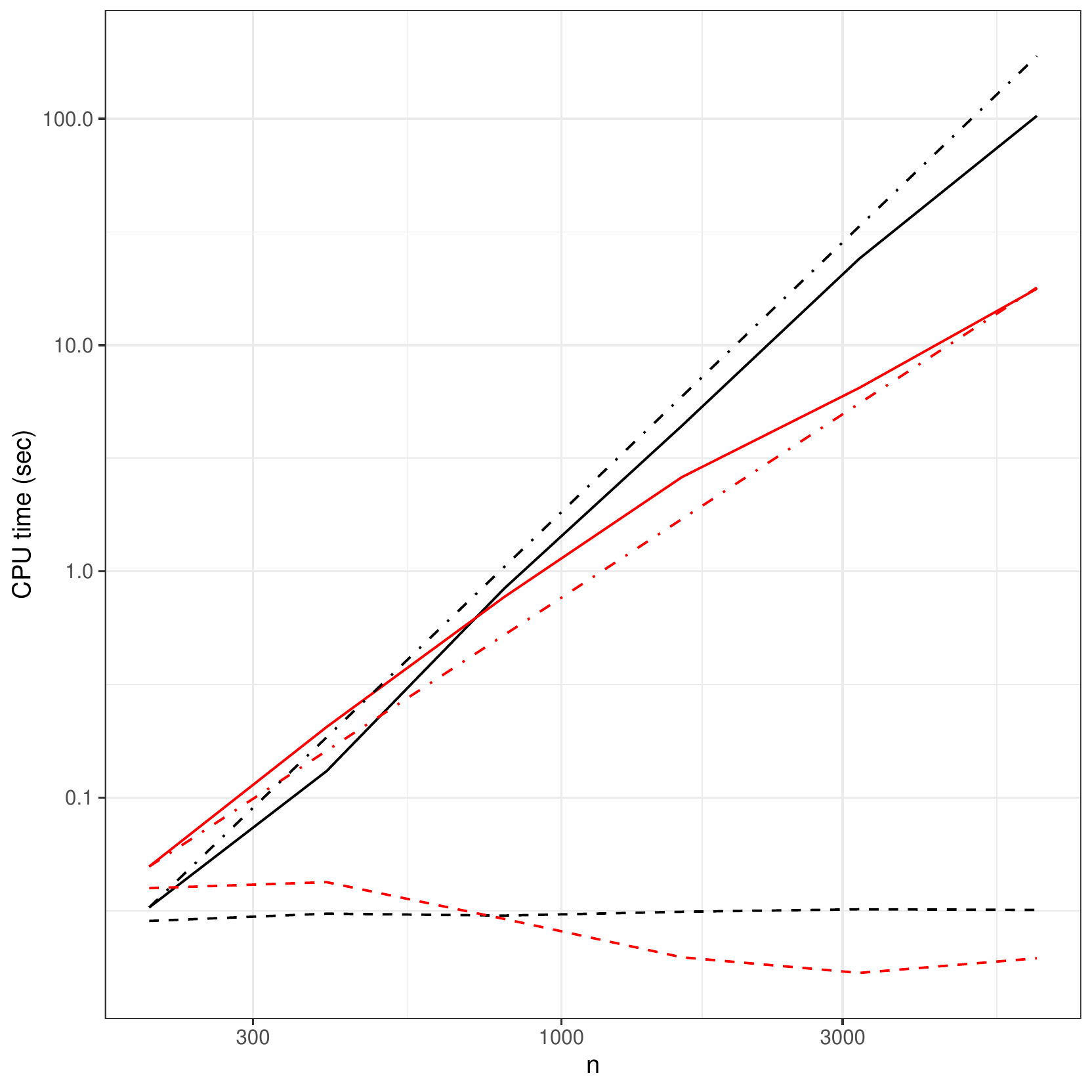}
\caption{Empirical computational cost for \code{cpop} as a function of sample size, $n$, for a grid of size $n$ (full lines) and of size $200$ (dashed lines), and for data with a single changepoint (black) and for a linearly increasing number of changepoints (red). To aid interpretation straight lines for CPU cost proportional to $n^{1.7}$ (red dot-dashed) and $n^{2.5}$ (black dot-dashed) are shown.}
\label{fig:cpop-CPU}
\end{figure}

The computational cost for \code{cpop} increases with the size of the number of potential changepoint locations. To see the rate of increase we ran \code{cpop} with the default settings where the grid of potential changepoints is equal to the $x$ values of the data, for data sizes varying from $n=200$ to $n=6400$. We considered two scenarios, one where we had a fixed number of changepoints and one where we had a fixed segment size of length 100. The average CPU cost across 10 runs of \code{cpop} for each data size are shown in Figure \ref{fig:cpop-CPU}. The suggest that the computational cost is increasing like $n^{2.5}$ when we have a fixed number of changes, and like $n^{1.7}$ when the number of changes increases linearly with $n$. By comparison, if we analyse each data set with a grid of 200 evenly spaced potential locations for the changes, the computational cost is roughly constant.

Thus for large data sets, we can substantially reduce the computational cost of running \code{cpop} by using a smaller grid of potential change locations. Obviously this comes with the drawback of a potential loss of accuracy with regards to the estimated changepoint locations. However one possible approach is to run \code{cpop} with a coarse grid, and then re-run the algorithm with a finer grid around the estimated changepoints.

To see this we implemented the scheme for a data set with $n=6400$ and a fixed segment size of 200.  We initially ran \code{cpop} with a grid with potential changes allowed every 16 observations.
\begin{CodeChunk}
\begin{CodeInput}
R> x <- 1:6400
R> y <- simchangeslope(x, changepoints = 0:31*200, 
+ change.slope = c(0.05,0.1*(-1)^(1:(31))), sigma = 1)
\end{CodeInput}
\end{CodeChunk}

We use a smaller value for the penalty due to the smaller grid size, and the fact that this is a preliminary step to find roughly where the changes are: so the key is to avoid missing changes. Spurious changes can still be removed when we perform our final run of \code{cpop}. 
\begin{CodeChunk}
\begin{CodeInput}
R> res.coarse <- cpop(y, x, grid=1:399*16, beta = 2 * log(400))
\end{CodeInput}
\end{CodeChunk}

In our example we find 38 changepoints with this coarse grid. We then introduce a finer grid around these putative changes: our new grid includes all $x$-values within 8 of each putative changepoint.
\begin{CodeChunk}
\begin{CodeInput}
R> cps <- unlist(changepoints(res.coarse))

R> grid <- NULL
R> for(i in 1:length(cps))
R> {
R>   grid <- c(grid, cps[i] + (-7):8)
R> }

R> res.fine <- cpop(y, x, grid, beta = 2 * log(length(x)))
\end{CodeInput}
\end{CodeChunk}

This gives a computational saving of between 10 to 100 over the default running of \code{cpop}. We can evaluate the accuracy of the approach by then comparing the estimated changepoints to the estimates we obtain if we run the default setting of \code{cpop}. In this case, both runs estimate the same number of changepoints, with the maximum difference in the location of a change being two time-points. The slower, default running of \code{cpop} gives a segmentation with a marginally lower cost (of 7187.1 as opposed to 7188.0).

\subsection{Imposing a Minimum Segment Length}

The \code{cpop} function allows the user to specify a minimum segment length -- and this is defined as the minimum $x$-distance allowed between two estimated changepoints. Specifying a minimum segment length can make the method more robust to point outliers or noise that is heavier-tailed than Gaussian: as minimising (\ref{eq:penalised_cost}) can lead to over-fitting in such scenarios and this over-fitting tends to be through adding clusters of changepoints close together to fit the noise in the data. There are two disadvantages of imposing a minimum segment length. First it can cause true changes to be missed if they are closer together than the specified minimum segment length. Second \code{cpop} is slower when a minimum segment length is imposed.

To see these issues, we simulated data as in Section \ref{sec:cpop}, except that we assumed the noise was $t_4$ distributed. We cannot simulate such data directly with \code{simchangeslope}, so we need to first use \code{simchangeslope} to calculate the mean function, and then add the noise:

\begin{CodeChunk}
\begin{CodeInput}
R> changepoints <- c(0, 25, 50, 100)
R> change.slope <- c(0.2, -0.3, 0.2, -0.1)
R> x <- 1:200
R> mu <- simchangeslope(x, changepoints, change.slope, sigma = 0)

R> y <- mu + rt(length(x), df = 4)
\end{CodeInput}
\end{CodeChunk}

We then estimated the changepoint locations both without a minimum segment length, and with minimum segment lengths of 10, 30 and 40. To run \code{cpop} with a minimum segment length of 10:
\begin{CodeChunk}
\begin{CodeInput}
R> res.min <- cpop( y, x, beta = 2*( log( length(y) ) ),
  +. minseglen = 10, sd = sqrt(2) )
\end{CodeInput}
\end{CodeChunk}
The argument \code{sd = sqrt(2)} is because $t_4$ distributed noise has a variance of 2.
%
%\begin{figure}
%\begin{minipage}{.5\linewidth}
%\centering
%{\includegraphics[scale=.5]{figures/cpop-minseg1.pdf}}
%\end{minipage}%
%\begin{minipage}{.5\linewidth}
%\centering
%{\includegraphics[scale=.5]{figures/cpop-minseg2.pdf}}
%\end{minipage}%
%\newline
%\begin{minipage}{.5\linewidth}
%\centering
%{\includegraphics[scale=.5]{figures/cpop-minseg3.pdf}}
%\end{minipage}%
%\begin{minipage}{.5\linewidth}
%\centering
%{\includegraphics[scale=.5]{figures/cpop-minseg4.pdf}}
%\end{minipage}%
%\centering
%\includegraphics{figures/cpop-example-uneven.pdf}
%\caption{Results of analysing data with $t_4$ noise with no minimum segment length (top left) and minimum segment lengths of 10 (top right), 30 (bottom left) and 40 (bottom right). In each plot we show data (grey dots), true mean (blue dashed line), true changepoints (blue vertical dashed lines), estimated mean (black line) and estimated changepoints (red vertical liens)}
%\label{fig:cpop-minseg}
%\end{figure}
%
%
\begin{figure}
\centering
{\includegraphics[scale=.5]{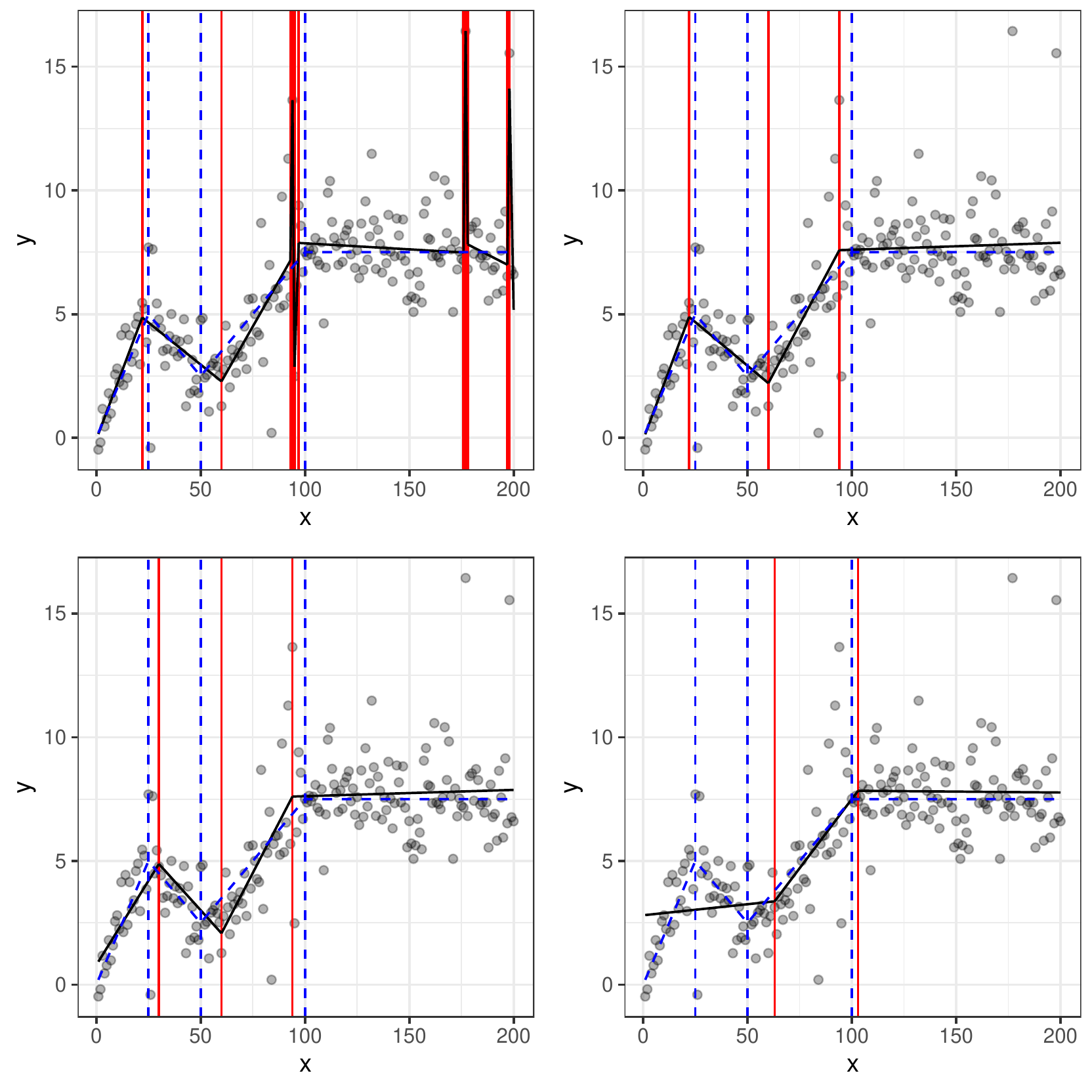}}
\caption{Results of analysing data with $t_4$ noise with no minimum segment length (top left) and minimum segment lengths of 10 (top right), 30 (bottom left) and 40 (bottom right). In each plot we show data (grey dots), true mean (blue dashed line), true changepoints (blue vertical dashed lines), estimated mean (black line) and estimated changepoints (red vertical liens)}
\label{fig:cpop-minseg}
\end{figure}
Results from CPOP with different minimum segment lengths are shown in Figure \ref{fig:cpop-minseg}. If we do not impose a minimum segment length, then we estimate 11 changepoints, including three cluster of changes that overfit to the noise. By imposing a minimum segment length of 10 we avoid the over-fitting. For this example, the computational cost of running \code{cpop} with the minimum segment length is about 15 times larger than when we do not assume a minimum segment length.

Assuming a minimum segment length of 30 or 40 shows what can happen when our minimum segment length assumption does not hold. A minimum segment length of 30 leads to estimates of the first two changes at time-points 30 and 60 -- the closest possible given the assumption. As we increase the minimum segment length to 40 we miss the first changepoint all together. 

\subsection{Choice of Penalty}

The choice of the penalty, \code{beta}, in \code{cpop} can have a substantial impact on the number of changepoints detected and the accuracy of the estimated mean function. This is common to all changepoint methods, where there will be at least one tuning parameter that specifies the evidence for a change that is needed before a change is added. The default choice of penalty, $2\log n$ where $n$ is the data size, is based on assumptions that the noise is IID Gaussian with known variance. When these assumptions do not hold, it is recommended to look at the segmentations obtained as the penalty value is varied: this can be done efficiently using the CROPS algorithm of \cite{haynes2017computationally}. 

The idea of CROPS is that it allows a penalised cost method to be implemented for all penalty values in an interval. This is implemented within the \pkg{cpop} package by the function:
\begin{CodeInput}
cpop.crops(y, x = 1:length(y), grid = x, beta_min = 1.5 * log(length(y)),
  beta_max = 2.5 * log(length(y)), sd = 1, minseglen = 0, prune.approx = FALSE)
\end{CodeInput}
The arguments of \code{cpop.crops} are identical to those of \code{cpop} except that, rather than specifying a single penalty value (\code{beta}), the range of penalty values to be used is specified by \code{beta_min} and \code{beta_max}, which fix the smallest and largest penalty value to be used. The output is an instance of an S4 class that contains details of all segmentations found by minimising the penalised cost for some penalty value in the interval between \code{beta_min} and \code{beta_max}.

To see the use of \code{cpop.crops}, consider an application where we do not know the standard deviation of the noise. Under our criteria (\ref{eq:penalised_cost}), the optimal segmentation with penalty $c^2 \beta$ and standard deviation $\sigma_i/c$ will be the same if we fix $\beta$ and $\sigma_{1:n}$ but vary $c>0$. Thus under an assumption that the noise is homogeneous, we can run \code{cpop} with \code{sd = 1} but for a range of $\beta \in [2\sigma_{-}^2\log n,2\sigma_{+}^2\log n]$, and this will give us the optimal segmentations for $\beta=2 \log n$ as we vary noise standard deviation between $\sigma_{-}$ and $\sigma_{+}$.

We simulated data as per Section \ref{sec:cpop} but with \code{sd = 1.5}. To run CPOP for a range of $\beta\in[7.5, 50]$ we use
\begin{CodeInput}
R> res.crops <- cpop.crops(y, x, beta_min = 5, beta_max = 50)
\end{CodeInput}
For our example $2 \log n = 10.6$, so this is equivalent to trying noise standard deviation in the range $[0.8,2.2]$. 

\begin{figure}
\centering
{\includegraphics[scale=.5,width=15cm,trim={0 4.4cm 0 4.4cm},clip]{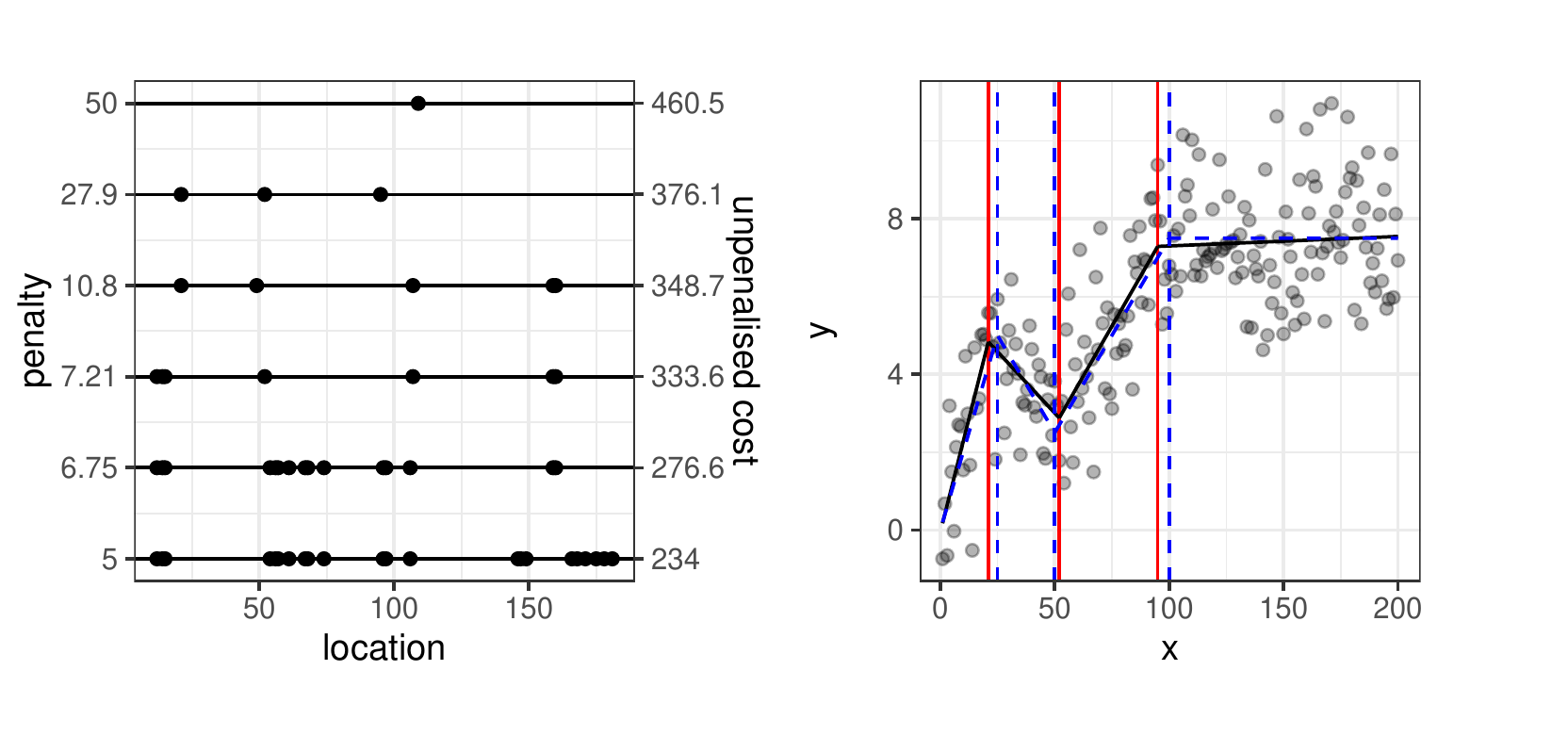}}
\caption{Example plot of output from \code{cpop.crops}
(left), and best segmentation based on calculated \code{BIC} (right). For the left-hand plot each row shows a segmentation, with points at estimated changepoint location. The left-hand axis shows a penalty value that leads to that segmentation, and the right-hand axis gives the corresponding unpenalised cost. For the right-hand plot we show the true mean and changepoints (blue dashed lines), and estimated mean (black line) and changepoints (red lines).}
\label{fig:crops.cpopA}
\end{figure}

We can plot the location in the changepoints for each segmentation found by CPOP for $\beta\in[7.5,50]$
\begin{CodeInput}
R> plot(res.crops)
\end{CodeInput}
This is shown in Figure \ref{fig:crops.cpopA}, and shows that there are 6 different segmentations found. These are labelled with a penalty value which gives that segmentation (left axis) and the unpenalised cost, i.e. the weighted RSS, for that segmentation and penalty value (right axis).

Details of the segmentations can be obtained using \code{segmentations(res.crops) }. This gives a matrix with one row for each of the segmentations, and each row contains a value of $\beta$ that gives that segmentation, the corresponding unpenalised cost, the penalised cost, the number of changepoints, and then the list of ordered changepoints. 
We can also obtain a list with the output from \code{cpop} corresponding to each segmentation, with \code{models( res.crops )}. For example, one approach to choose a segmentation from the output from \code{cpop.crops} is to find the segmentation that minimises a cost under a model where we assume the noise variance is unknown \citep{fryzlewicz2014wild},
\[
 n \log \left( \frac{1}{n} \sum_{i=1}^n \left(y_i-\hat{f}(x_i)\right)^2
\right) + 2 K \log n.
\]
Here $\hat{f}$ is the estimated mean function and $K$ is the number of changepoints. This can be calculated as follows.
\begin{CodeChunk}
\begin{CodeInput}
R> models <- cpop.crops.models(res.crops)
R> M <- length(models)
R> BIC <- rep(NA, M)
R> ncps <- segmentations(res.crops )[,4]
R> n <- length(y)
R> for(j in 1:M)
R> {
R>   BIC[i] <- n * log(mean((residuals(models[[j]]))^2)) 
+    + 2 * ncps[i] * log(n)
R> }
\end{CodeInput}
\end{CodeChunk}
This uses that the fourth column of the matrix \code{segmentations( res.crops )} stores the number of changepoints in each segmentation, and that we can calculate $y_i - \hat{f}(x_i)$ using the \code{residual} function evaluated for the corresponding entry of \code{cpop.crops.models(res.crops)}. The segmentation which has the smallest value of \code{BIC} is shown in Figure \ref{fig:crops.cpopA}, and shows that this correctly chooses the segmentation with three changes.

%\begin{figure}
%\begin{minipage}{.5\linewidth}
%\centering
%{\includegraphics[scale=.55]{figures/cpop-cropsA1.pdf}}
%\end{minipage}%
%\begin{minipage}{.5\linewidth}
%\centering
%{\includegraphics[scale=.55]{figures/cpop-cropsA2.pdf}}
%\end{minipage}%
%\caption{Example plot of output from \code{cpop.crops}
%(left), and best segmentation based on calculated \code{BIC} %(right). For the left-hand plot each row shows a %segmentation, with points at estimated changepoint location. %The left-hand axis shows a penalty value that leads to that %segmentation, and the right-hand axis gives the %corresponding unpenalised cost. For the right-hand plot we %show the true mean and changepoints (blue dashed lines), and %estimated mean (black line) and changepoints (red lines).}
%\label{fig:crops.cpopA}
%\end{figure}
%
%
As a final example, we performed a similar analysis but with correlated noise. This violates the assumption of IID noise that underpins the default choice of penalty, thus we run \code{cpop.crops} for a range of penalties. 

We simulated data with $n=500$ data points and $10$ equally spaced changepoints.
\begin{CodeChunk}
\begin{CodeInput}
R> n <- 500
R> x <- 1:n
R> mu <- simchangeslope(x, changepoints = 45*0:10, 
+  change.slope = c(0.15,0.3*(-1)^(1:10)), sd = 0)
R> epsilon <- rnorm(n+2)
R> y <- mu + (epsilon[1:n] + epsilon[2:(n+1)] + epsilon[3:(n+2)]) / sqrt(3)
\end{CodeInput}
\end{CodeChunk}
The noise is MA(3), and we simulate the data by first calculating the mean function, \code{mu}, and then adding the MA(3) noise.

We could continue as above, and choose between the segmentations by minimising a penalised cost under an appropriate model for the noise; this type of approach is suggested for change in mean models by \cite{cho2020multiple}. A simpler, albeit more qualitative approach, is to plot the residual sum of squares of the segmentation against the number of changepoints \citep{lebarbier2005detecting,baudry2012slope,fearnhead2020relating,fryzlewicz2020detecting}. This avoids the need to specify a model for the residuals. The idea of this approach is that adding ``true" changes should lead to a noticeably larger reduction in the residual sum of squares than adding ``spurious" changes. Thus the best segmentation should correspond to an ``elbow" in this plot.

\begin{figure}
\centering
{\includegraphics{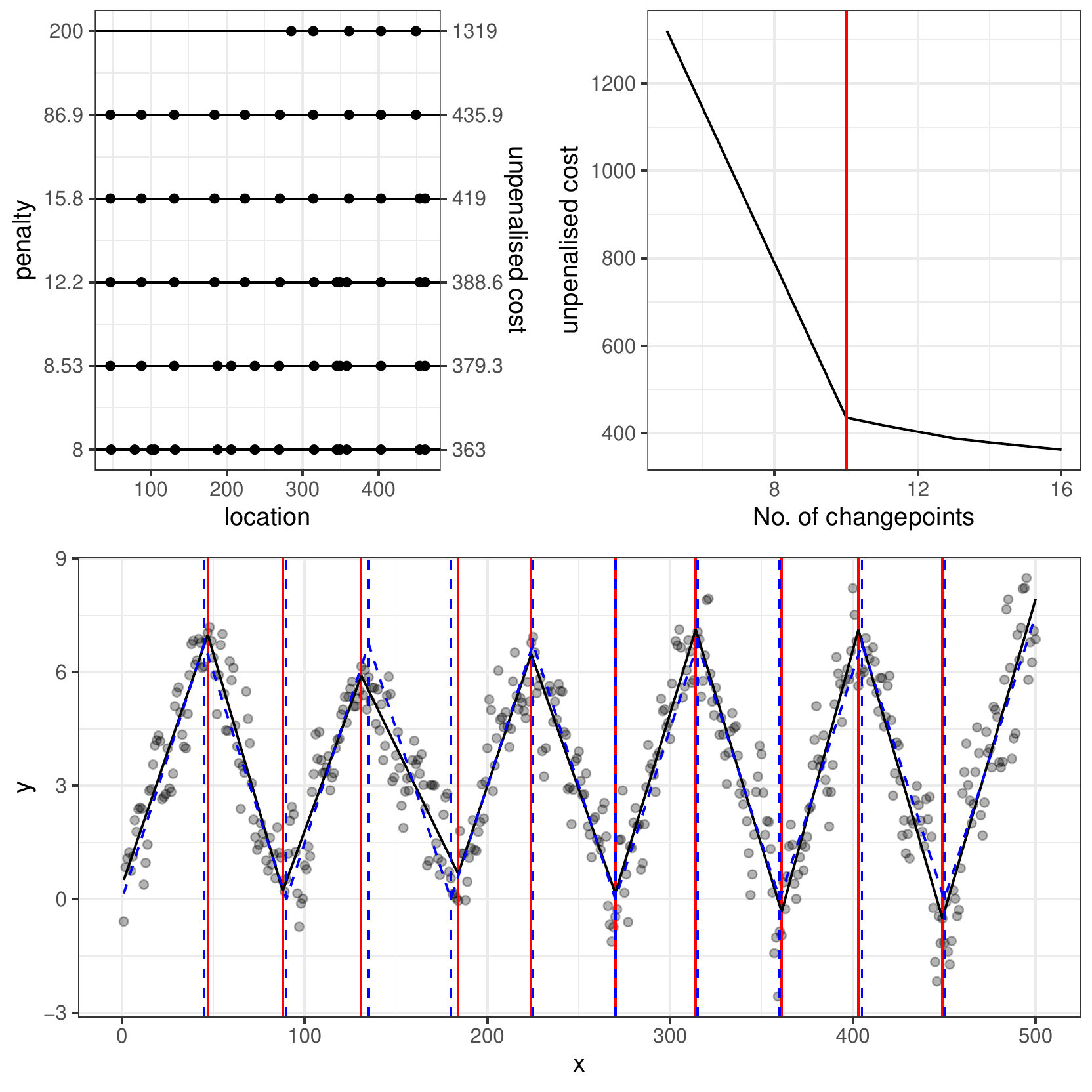}}
\caption{Correlated noise example. Output from \code{cpop.crops} (top left), unpenalised cost against number of changepoint (top right) and estimate from segmentation corresponding to "elbow" (bottom).}
\label{fig:crops.cpopB}
\end{figure}

\begin{CodeChunk}
\begin{CodeInput}
R> res.crops <- cpop.crops(y, x, beta_min = 8, beta_max = 200)
R> segs <- segmentations(res.crops)
R> p <- ggplot(data = segs, aes(x = m))
R> p <- p + geom_line(aes(y = Qm))
R> p <- p + geom_vline(xintercept = 10, color = "red")
R> p <- p + xlab("No. of changepoints") + ylab("unpenalised cost")
R> plot(p)
\end{CodeInput}
\end{CodeChunk}
This runs \code{cpop.crops} and then uses the fact that the output of \code{segmentations} includes columns that give the number of changepoints and the unpenalised cost of each segmentation. These columns are labelled \code{"m"} and \code{"Qm"} respectively.
The plot gives a clear elbow, see Figure \ref{fig:crops.cpopB}, and this corresponds to a correct estimate of the number of changes. %The fitted changes and estimated mean function are also shown in Figure \ref{fig:crops.cpopB}.
%
%\begin{figure}
%\begin{minipage}{.5\linewidth}
%\centering
%{\includegraphics[scale=.5]{figures/cpop-cropsB1.pdf}}
%\end{minipage}%
%\begin{minipage}{.5\linewidth}
%\centering
%{\includegraphics[scale=.5]{figures/cpop-cropsB2.pdf}}
%\end{minipage}%
%\newline
%\begin{minipage}{1.0\linewidth}
%\centering
%{\includegraphics[scale=.5]{figures/cpop-cropsB3.pdf}}
%\end{minipage}%
%\caption{Correlated noise example. Output from \code{cpop.crops} (top left), unpenalised cost against number of changepoint (rop right) and estimate from segmentation corresponding to "elbow" (bottom).}
%\label{fig:crops.cpopB}
%\end{figure}
%

\section{Application} \label{sec:app}

\begin{figure}
\centering
{\includegraphics{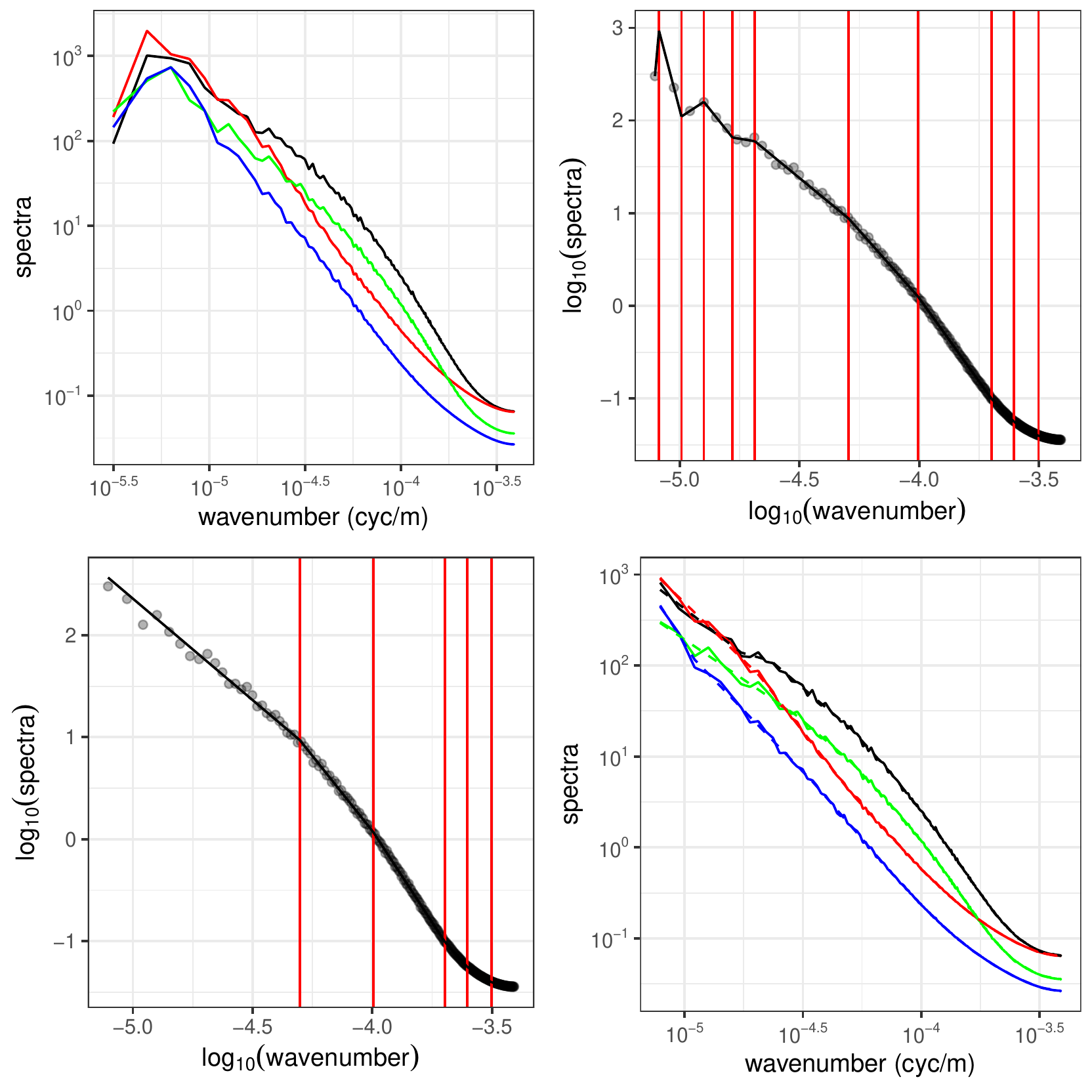}}
\caption{Application of \code{cpop} to \code{wavenumber\_spectra} data. Log-log plot of raw data (top left) of horizontal wavenumber spectra of velocity for two months and two runs of an ocean model: February 2000 (black), August 2000 (red), February 2100 (green) and August 2100 (blue). Output from \code{cpop} applied to analyse the decay of spectra from August 2000, with $y$ equal to log spectra and $x$ equal to log wave number, with estimated homogeneous noise variance (top right) and estimated heterogeneous noise variance (bottom left). Log-log plot of fitted spectra for all four series (bottom right) with original data in full-lines and estimate in dashed lines. 
}
\label{fig:crops.data}
\end{figure}

We now demonstrate an application of \code{cpop} on analysing power spectra of velocity as a function of wavenumber obtained from models of the Atlantic Ocean. The data is available in the \pkg{cpop} package and can be loaded with \code{data("wavenumber_spectra")}. It contains four spectra, corresponding to two different months (February and August) from two different runs of the model (2000 and 2100) corresponding to present and future scenarios: see Figure \ref{fig:crops.data}. The data comes from \cite{richards2021impact}, and is available from \cite{richards2020data}. See \cite{richards2021impact} for a fuller description of the data.

Interest lies in estimating the rate of decay of the log-spectra against log-wavenumber. We can do this by removing the first three data points (where the spectra is increasing) and then using \code{cpop} to fit a piecewise-linear curve to the remaining data. We perform an initial run of \code{cpop} assuming an estimated homogeneous noise variance on the data from August from the 2000 run.
\begin{CodeChunk}
\begin{CodeInput}
R> data("wavenumber_spectra")
R> x <- log(wavenumber_spectra[-(1:3),1])
R> y <- log(wavenumber_spectra[-(1:3),4])

R> grid <- seq(from = min(x), to = max(x), length = 200)

R> sig2 <- mean( diff( diff(y) )^2 )/6 

R> res <- cpop(y, x, grid, sd = sqrt(sig2), minseg = 0.2, beta = 2*log(200))
\end{CodeInput}
\end{CodeChunk}
Here we estimate the noise variance, \code{sig2}, based on the variance of the double difference of the data. For regions where the mean is linear and the data is evenly spaced, taking the double difference will lead to a mean zero process. If the noise is IID with variance $\sigma^2$ then the double-differenced process will have a marginal variance that is $6\sigma^2$. Thus our estimate is the empirical mean of the square of the double-difference data divided by 6. The original data is evenly spaced in in terms of wavenumber, but as we take logs, \code{x} is unevenly spaced: so this estimator will be biased in our setting. However it will give a reasonable ball-park figure for an initial run of \code{cpop} -- the residuals from which can then be used to get a better estimate of the noise variance. We use a evenly spaced grid for possible change-point locations. To avoid the potential for adding multiple changepoints between two observations, we set a minimum segment length of 0.09 (as the largest distance between consecutive \code{x} values is 0.08).

The output is shown in the top-right plot of Figure \ref{fig:crops.data}, and appears to be over-fitting to the early part of the series. This is because the noise variance is heterogeneous, and decreasing with \code{x}. However, given our initial fit we can use the residuals to estimate the noise variance. The noise for the spectra is expected to be approximately inversely proportional to the wavenumber. By using a Taylor-expansion, we have the variance of the noise for the log-spectra should be approximately the variance of the noise of the spectra divided by the square of the mean of the spectra. As the mean of the spectra is roughly a power of the wavenumber, this suggests using a model for the variance, $\sigma^2_x$ say, depending on $x$ as $\log \sigma^2_x = a + bx$: so that the variance is proportional to some power of the wavenumber. We can estimate the parameters of this model by maximising the log-likelihood of Gaussian model for the residuals with this form for the variance.
\begin{CodeChunk}
\begin{CodeInput}
R> r2 <- residuals(res)^2
R> loglik <- function(par)
R> { 
R>  return(length(r2) * par[1] + par[2] * sum(x) 
  + sum(r2 / (exp(par[1]+par[2]*x))))
R> }
R> est.hat <- optim(c(0,0), loglik)
R> sig2 <- exp(est.hat$par[1] + est.hat$par[2]*x)
R> res2 <- cpop(y, x, grid, sd = sqrt(sig2), 
+               minseg = 0.2, beta =2 * log(200))
\end{CodeInput}
\end{CodeChunk}
Here we have calculated the maximum likelihood estimates by using \code{optim} to minimise minus the log-likelihood. The resulting output from \code{cpop} is shown in the bottom left plot of Figure \ref{fig:crops.data}.  The first two changes could represent real regime transitions relating to the inviscid fluid physics that one would see in the real ocean \cite[see Figure 6a of][]{callies2013interpreting}, while the three changes for the largest values of \code{x} may relate to a breakdown in the numerical ocean model near the highest wavenumber of the ocean model grid \cite[]{soufflet2016effective}. The estimates of the spectra for all four series, obtained by repeating this approach, is also shown in Figure \ref{fig:crops.data}. For this application, the residuals from the fitted model appear to be uncorrelated and sub-Gaussian, so using the fit based on the default penalty choice is reasonable. Though one could also explore segmentations for other penalty choices using \code{cpop.crops} as in the previous section.

\section{Acknowledgements}
We would like to thanks Jessica Luo and Dan Whitt for access to and help with the wavenumber spectra data. Paul Fearnhead acknowledges funding from EPSRC grant EP/N031938/1.

%\section{Discussion}

%\bibliographystyle{jss}
\bibliography{refs}

\begin{thebibliography}{43}
\newcommand{\enquote}[1]{``#1''}
\providecommand{\natexlab}[1]{#1}
\providecommand{\url}[1]{\texttt{#1}}
\providecommand{\urlprefix}{URL }
\expandafter\ifx\csname urlstyle\endcsname\relax
  \providecommand{\doi}[1]{doi:\discretionary{}{}{}#1}\else
  \providecommand{\doi}{doi:\discretionary{}{}{}\begingroup
  \urlstyle{rm}\Url}\fi
\providecommand{\eprint}[2][]{\url{#2}}

\bibitem[{Aminikhanghahi and Cook(2017)}]{aminikhanghahi2017survey}
Aminikhanghahi S, Cook DJ (2017).
\newblock \enquote{A Survey of Methods for Time Series Change Point Detection.}
\newblock \emph{Knowledge and Information Systems}, \textbf{51}(2), 339--367.
\newblock \doi{10.1007/s10115-016-0987-z}.

\bibitem[{Anastasiou and Fryzlewicz(2022)}]{anastasiou2022detecting}
Anastasiou A, Fryzlewicz P (2022).
\newblock \enquote{Detecting Multiple Generalized Change-points by Isolating
  Single Ones.}
\newblock \emph{Metrika}, \textbf{85}(2), 141--174.
\newblock \doi{10.1007/s00184-021-00821-6}.

\bibitem[{Andreou and Ghysels(2002)}]{andreou2002detecting}
Andreou E, Ghysels E (2002).
\newblock \enquote{Detecting Multiple Breaks in Financial Market Volatility
  Dynamics.}
\newblock \emph{Journal of Applied Econometrics}, \textbf{17}(5), 579--600.
\newblock \doi{10.1002/jae.684}.

\bibitem[{{Arnold, Taylor B and Tibshirani, Ryan J}(2020)}]{genlasso}
{Arnold, Taylor B and Tibshirani, Ryan J} (2020).
\newblock \emph{\pkg{genlasso}: Path Algorithm for Generalized Lasso Problems}.
\newblock \urlprefix\url{https://github.com/glmgen/genlasso}.

\bibitem[{Baranowski \emph{et~al.}(2019)Baranowski, Chen, and
  Fryzlewicz}]{baranowski2016narrowest}
Baranowski R, Chen Y, Fryzlewicz P (2019).
\newblock \enquote{Narrowest-Over-Threshold Detection of Multiple Change-Points
  and Change-Point-like Features.}
\newblock \emph{Journal of the Royal Statistical Society B}, \textbf{81},
  649--672.
\newblock \doi{10.1111/rssb.12322}.

\bibitem[{Baudry \emph{et~al.}(2012)Baudry, Maugis, and
  Michel}]{baudry2012slope}
Baudry JP, Maugis C, Michel B (2012).
\newblock \enquote{Slope Heuristics: Overview and Implementation.}
\newblock \emph{Statistics and Computing}, \textbf{22}(2), 455--470.
\newblock \doi{10.1007/s11222-011-9236-1}.

\bibitem[{Callies and Ferrari(2013)}]{callies2013interpreting}
Callies J, Ferrari R (2013).
\newblock \enquote{Interpreting Energy and Tracer Spectra of Upper-Ocean
  Turbulence in the Submesoscale range (1--200 km).}
\newblock \emph{Journal of Physical Oceanography}, \textbf{43}(11), 2456--2474.
\newblock \doi{10.1175/JPO-D-13-063.1}.

\bibitem[{Cho and Fryzlewicz(2020)}]{cho2020multiple}
Cho H, Fryzlewicz P (2020).
\newblock \enquote{Multiple Change Point Detection under Serial Dependence:
  Wild Energy Maximisation and Gappy Schwarz Criterion.}
\newblock \emph{arXiv preprint arXiv:2011.13884}.

\bibitem[{Erdman and Emerson(2008)}]{erdman2008bcp}
Erdman C, Emerson JW (2008).
\newblock \enquote{\pkg{bcp}: an \proglang{R} Package for Performing a Bayesian
  Analysis of Change Point Problems.}
\newblock \emph{Journal of Statistical Software}, \textbf{23}, 1--13.
\newblock \doi{10.18637/jss.v023.i03}.

\bibitem[{Fearnhead \emph{et~al.}(2019)Fearnhead, Maidstone, and
  Letchford}]{fearnhead2019detecting}
Fearnhead P, Maidstone R, Letchford A (2019).
\newblock \enquote{Detecting Changes in Slope with an $l_0$ Penalty.}
\newblock \emph{Journal of Computational and Graphical Statistics},
  \textbf{28}(2), 265--275.
\newblock \doi{10.1080/10618600.2018.1512868}.

\bibitem[{Fearnhead and Rigaill(2020)}]{fearnhead2020relating}
Fearnhead P, Rigaill G (2020).
\newblock \enquote{Relating and Comparing Methods for Detecting Changes in
  Mean.}
\newblock \emph{Stat}, \textbf{9}(1), e291.
\newblock \doi{10.1002/sta4.291}.

\bibitem[{Frick \emph{et~al.}(2014)Frick, Munk, and
  Sieling}]{frick2014multiscale}
Frick K, Munk A, Sieling H (2014).
\newblock \enquote{Multiscale Change Point Inference.}
\newblock \emph{Journal of the Royal Statistical Society B}, \textbf{76}(3),
  495--580.
\newblock \doi{10.1111/rssb.12047}.

\bibitem[{Fryzlewicz(2014)}]{fryzlewicz2014wild}
Fryzlewicz P (2014).
\newblock \enquote{Wild Binary Segmentation for Multiple Change-point
  Detection.}
\newblock \emph{The Annals of Statistics}, \textbf{42}(6), 2243--2281.
\newblock \doi{10.1214/14-AOS1245}.

\bibitem[{Fryzlewicz(2020)}]{fryzlewicz2020detecting}
Fryzlewicz P (2020).
\newblock \enquote{Detecting Possibly Frequent Change-points: Wild Binary
  Segmentation 2 and Steepest-drop Model Selection.}
\newblock \emph{Journal of the Korean Statistical Society}, \textbf{49}(4),
  1027--1070.
\newblock \doi{10.1007/s42952-020-00060-x}.

\bibitem[{Grundy \emph{et~al.}(2020)Grundy, Killick, and
  Mihaylov}]{grundy2020high}
Grundy T, Killick R, Mihaylov G (2020).
\newblock \enquote{High-dimensional Changepoint Detection via a Geometrically
  Inspired Mapping.}
\newblock \emph{Statistics and Computing}, \textbf{30}(4), 1155--1166.
\newblock \doi{10.1007/s11222-020-09940-y}.

\bibitem[{Haynes \emph{et~al.}(2017{\natexlab{a}})Haynes, Eckley, and
  Fearnhead}]{haynes2017computationally}
Haynes K, Eckley IA, Fearnhead P (2017{\natexlab{a}}).
\newblock \enquote{Computationally Efficient Changepoint Detection for a Range
  of Penalties.}
\newblock \emph{Journal of Computational and Graphical Statistics},
  \textbf{26}(1), 134--143.
\newblock \doi{10.1080/10618600.2015.1116445}.

\bibitem[{Haynes \emph{et~al.}(2017{\natexlab{b}})Haynes, Fearnhead, and
  Eckley}]{haynes2017NP}
Haynes K, Fearnhead P, Eckley IA (2017{\natexlab{b}}).
\newblock \enquote{A Computationally Efficient Nonparametric Approach for
  Changepoint Detection.}
\newblock \emph{Statistics and Computing}, \textbf{27}(5), 1293--1305.
\newblock \doi{10.1007/s11222-016-9687-5}.

\bibitem[{Jackson \emph{et~al.}(2005)Jackson, Scargle, Barnes, Arabhi, Alt,
  Gioumousis, Gwin, Sangtrakulcharoen, Tan, and Tsai}]{jackson2005algorithm}
Jackson B, Scargle JD, Barnes D, Arabhi S, Alt A, Gioumousis P, Gwin E,
  Sangtrakulcharoen P, Tan L, Tsai TT (2005).
\newblock \enquote{An Algorithm for Optimal Partitioning of Data on an
  Interval.}
\newblock \emph{IEEE Signal Processing Letters}, \textbf{12}(2), 105--108.
\newblock \doi{10.1109/LSP.2001.838216}.

\bibitem[{James \emph{et~al.}(2015)James, Matteson
  \emph{et~al.}}]{james2015ecp}
James NA, Matteson DS, \emph{et~al.} (2015).
\newblock \enquote{\pkg{ecp}: {An \proglang{R}} Package for Nonparametric
  Multiple Change Point Analysis of Multivariate Data.}
\newblock \emph{Journal of Statistical Software}, \textbf{62}(i07), 1--25.
\newblock \doi{10.18637/jss.v062.i07}.

\bibitem[{Jewell \emph{et~al.}(2020)Jewell, Hocking, Fearnhead, and
  Witten}]{jewell2020fast}
Jewell SW, Hocking TD, Fearnhead P, Witten DM (2020).
\newblock \enquote{Fast Nonconvex Deconvolution of Calcium Imaging Data.}
\newblock \emph{Biostatistics}, \textbf{21}(4), 709--726.
\newblock \doi{10.1093/biostatistics/kxy083}.

\bibitem[{Killick and Eckley(2014)}]{killick2014changepoint}
Killick R, Eckley I (2014).
\newblock \enquote{\pkg{changepoint}: {An \proglang{R}} Package for Changepoint
  Analysis.}
\newblock \emph{Journal of Statistical Software}, \textbf{58}(3), 1--19.
\newblock \doi{10.18637/jss.v058.i03}.

\bibitem[{Killick \emph{et~al.}(2012)Killick, Fearnhead, and
  Eckley}]{killick2012optimal}
Killick R, Fearnhead P, Eckley IA (2012).
\newblock \enquote{Optimal Detection of Changepoints with a Linear
  Computational Cost.}
\newblock \emph{Journal of the American Statistical Association},
  \textbf{107}(500), 1590--1598.
\newblock \doi{10.1080/01621459.2012.737745}.

\bibitem[{Kim \emph{et~al.}(2009)Kim, Koh, Boyd, and Gorinevsky}]{kim2009ell_1}
Kim SJ, Koh K, Boyd S, Gorinevsky D (2009).
\newblock \enquote{$\ell_1$ Trend Filtering.}
\newblock \emph{SIAM Review}, \textbf{51}(2), 339--360.
\newblock \doi{10.1137/070690274}.

\bibitem[{Kov{\'a}cs \emph{et~al.}(2022)Kov{\'a}cs, Li, B{\"u}hlmann, and
  Munk}]{kovacs2020seeded}
Kov{\'a}cs S, Li H, B{\"u}hlmann P, Munk A (2022).
\newblock \enquote{Seeded Binary Segmentation: {A} General Methodology for Fast
  and Optimal Change Point Detection.}
\newblock \emph{Biometrika, to appear}.

\bibitem[{Lebarbier(2005)}]{lebarbier2005detecting}
Lebarbier {\'E} (2005).
\newblock \enquote{Detecting Multiple Change-points in the Mean of Gaussian
  Process by Model Selection.}
\newblock \emph{Signal processing}, \textbf{85}(4), 717--736.
\newblock \doi{10.1016/j.sigpro.2004.11.012}.

\bibitem[{Li \emph{et~al.}(2016)Li, Munk, and Sieling}]{li2016fdr}
Li H, Munk A, Sieling H (2016).
\newblock \enquote{{FDR}-control in Multiscale Change-point Segmentation.}
\newblock \emph{Electronic Journal of Statistics}, \textbf{10}(1), 918--959.
\newblock \doi{10.1214/16-EJS1131}.

\bibitem[{Maidstone \emph{et~al.}(2017)Maidstone, Hocking, Rigaill, and
  Fearnhead}]{maidstone2017optimal}
Maidstone R, Hocking T, Rigaill G, Fearnhead P (2017).
\newblock \enquote{On Optimal Multiple Changepoint Algorithms for Large Data.}
\newblock \emph{Statistics and Computing}, \textbf{27}(2), 519--533.
\newblock \doi{10.1007/s11222-016-9636-3}.

\bibitem[{Matteson and James(2014)}]{matteson2014nonparametric}
Matteson DS, James NA (2014).
\newblock \enquote{A Nonparametric Approach for Multiple Change Point Analysis
  of Multivariate Data.}
\newblock \emph{Journal of the American Statistical Association},
  \textbf{109}(505), 334--345.
\newblock \doi{10.1080/01621459.2013.849605}.

\bibitem[{Meier \emph{et~al.}(2021)Meier, Kirch, and Cho}]{meier2021mosum}
Meier A, Kirch C, Cho H (2021).
\newblock \enquote{\pkg{mosum}: A Package for Moving Sums in Change-point
  Analysis.}
\newblock \emph{Journal of Statistical Software}, \textbf{97}, 1--42.
\newblock \doi{10.18637/jss.v097.i08}.

\bibitem[{Niu and Zhang(2012)}]{niu2012screening}
Niu YS, Zhang H (2012).
\newblock \enquote{{The Screening and Ranking Algorithm to Detect DNA Copy
  Number Variations}.}
\newblock \emph{The Annals of Applied Statistics}, \textbf{6}(3), 1306 -- 1326.
\newblock \doi{10.1214/12-AOAS539}.

\bibitem[{Pein \emph{et~al.}(2017)Pein, Sieling, and
  Munk}]{pein2017heterogeneous}
Pein F, Sieling H, Munk A (2017).
\newblock \enquote{Heterogeneous Change Point Inference.}
\newblock \emph{Journal of the Royal Statistical Society B}, \textbf{79}(4),
  1207--1227.
\newblock \doi{10.1111/rssb.12202}.

\bibitem[{Reeves \emph{et~al.}(2007)Reeves, Chen, Wang, Lund, and
  Lu}]{reeves2007review}
Reeves J, Chen J, Wang XL, Lund R, Lu QQ (2007).
\newblock \enquote{A Review and Comparison of Changepoint Detection Techniques
  for Climate Data.}
\newblock \emph{Journal of applied meteorology and climatology},
  \textbf{46}(6), 900--915.
\newblock \doi{10.1175/JAM2493.1}.

\bibitem[{Richards \emph{et~al.}(2020)Richards, Whitt, Brett, Bryan, Feloy, and
  Long}]{richards2020data}
Richards KJ, Whitt DB, Brett G, Bryan FO, Feloy K, Long MC (2020).
\newblock \emph{Climate Change Impact on Submesoscale ROMS Data}.
\newblock \urlprefix\url{https://zenodo.org/record/4615129#.Ysazwi8w2X0}.

\bibitem[{Richards \emph{et~al.}(2021)Richards, Whitt, Brett, Bryan, Feloy, and
  Long}]{richards2021impact}
Richards KJ, Whitt DB, Brett G, Bryan FO, Feloy K, Long MC (2021).
\newblock \enquote{{The Impact of Climate Change on Ocean Submesoscale
  Activity}.}
\newblock \emph{Journal of Geophysical Research: Oceans}, \textbf{126}(5),
  e2020JC016750.
\newblock \doi{10.1029/2020JC016750}.

\bibitem[{Romano \emph{et~al.}(2021)Romano, Eckley, Fearnhead, and
  Rigaill}]{romano2021fast}
Romano G, Eckley I, Fearnhead P, Rigaill G (2021).
\newblock \enquote{Fast Online Changepoint Detection via Functional Pruning
  CUSUM Statistics.}
\newblock \emph{arXiv preprint arXiv:2110.08205}.

\bibitem[{Ross(2015)}]{ross2015parametric}
Ross GJ (2015).
\newblock \enquote{Parametric and Nonparametric Sequential Change Detection in
  \proglang{R}: {T}he \pkg{cpm} Package.}
\newblock \emph{Journal of Statistical Software}, \textbf{66}, 1--20.
\newblock \doi{10.18637/jss.v066.i03}.

\bibitem[{Runge \emph{et~al.}(2020)Runge, Hocking, Romano, Afghah, Fearnhead,
  and Rigaill}]{runge2020gfpop}
Runge V, Hocking TD, Romano G, Afghah F, Fearnhead P, Rigaill G (2020).
\newblock \enquote{\pkg{gfpop}: an \proglang{R} Package for Univariate
  Graph-constrained Change-point Detection.}
\newblock \emph{arXiv preprint arXiv:2002.03646}.

\bibitem[{Scott and Knott(1974)}]{scott1974cluster}
Scott AJ, Knott M (1974).
\newblock \enquote{A Cluster Analysis Method for Grouping Means in the Analysis
  of Variance.}
\newblock \emph{Biometrics}, \textbf{30}(3), 507--512.

\bibitem[{Shi \emph{et~al.}(2022)Shi, Gallagher, Lund, and
  Killick}]{shi2022comparison}
Shi X, Gallagher C, Lund R, Killick R (2022).
\newblock \enquote{A Comparison of Single and Multiple Changepoint Techniques
  for Time Series Data.}
\newblock \emph{Computational Statistics \& Data Analysis}, \textbf{170},
  107433.
\newblock \doi{10.1016/j.csda.2022.107433}.

\bibitem[{Soufflet \emph{et~al.}(2016)Soufflet, Marchesiello, Lemari{\'e},
  Jouanno, Capet, Debreu, and Benshila}]{soufflet2016effective}
Soufflet Y, Marchesiello P, Lemari{\'e} F, Jouanno J, Capet X, Debreu L,
  Benshila R (2016).
\newblock \enquote{On effective resolution in ocean models.}
\newblock \emph{Ocean Modelling}, \textbf{98}, 36--50.
\newblock \doi{10.1016/j.ocemod.2015.12.004}.

\bibitem[{Tibshirani(2014)}]{tibshirani2014adaptive}
Tibshirani RJ (2014).
\newblock \enquote{Adaptive Piecewise Polynomial Estimation via Trend
  Filtering.}
\newblock \emph{The Annals of Statistics}, \textbf{42}(1), 285--323.
\newblock \doi{10.1214/13-AOS1189}.

\bibitem[{Truong \emph{et~al.}(2020)Truong, Oudre, and
  Vayatis}]{truong2020selective}
Truong C, Oudre L, Vayatis N (2020).
\newblock \enquote{Selective Review of Offline Change Point Detection Methods.}
\newblock \emph{Signal Processing}, \textbf{167}, 107299.
\newblock \doi{10.1016/j.sigpro.2019.107299}.

\bibitem[{Wang and Samworth(2018)}]{wang2018high}
Wang T, Samworth RJ (2018).
\newblock \enquote{High Dimensional Change Point Estimation via Sparse
  Projection.}
\newblock \emph{Journal of the Royal Statistical Society B}, \textbf{80}(1),
  57--83.
\newblock \doi{10.1111/rssb.12243}.

\end{thebibliography}

\appendix

\section{Details of the Recursion} \label{App:cost_update}

Here we describe how to calculate the inner minimisation
\[
\min_{\alpha'} \left[
F_{k}(\alpha')+\mathcal{C}_{l-1,l}(\alpha',\alpha) +\beta
\right]
\]
in the dynamic programming recursion -- and how to do this so that the computational cost does not increase with the number of observations since the putative most recent changepoint.

The function $F_k(\alpha')$ will be defined as the minimum of a set of quadratics. Denote this $q^{(k)}_i(\alpha')$ for $i=1,\ldots,M_k$. Then we wish to solve
\[
\min_{\alpha'} \left[ \min_{i\in1:M_k} \left\{
q_{i}^{(k)}(\alpha')+\mathcal{C}_{k,l}(\alpha',\alpha)\right\} +\beta
\right] =
\min_{i\in 1:M_k}\left\{
\min_{\alpha'} \left[
q_{i}^{(k)}(\alpha')+\mathcal{C}_{k,l}(\alpha',\alpha) + \beta
\right]
\right\}.
\]
So we only need to be able to calculate $\min_{\alpha'} [q(\alpha')+\mathcal{C}_{l-1,l}(\alpha',\alpha)]$, for any known quadratic $q(\alpha')$. In the following, we will denote the co-coefficients of $q(\alpha')$ by $a$, $b$ and $c$, so
\[
q(\alpha')=a+b\alpha' + c\alpha'^2.
\]

Our approach will be to (i)  calculate the co-coefficients of $\mathcal{C}_{k,l}(\alpha',\alpha)$ in constant time, through the use of summary statistics; (ii) calculate the co-coefficients of the sum $q_{i}^{(k)}(\alpha')+\mathcal{C}_{k,l}(\alpha',\alpha) + \beta$; (iii) calculate the co-coefficients of the quadratic in $\alpha$ after we minimise with respect to $\alpha'$. Steps (ii) and (iii) are trivial, but we give details below for completeness.

To simplify the exposition in the following we will use the convention that expressions that are of the form $0/0$ are equal to 0.

Define the following summary statistics for the data. These can be calculated prior to solving the dynamic programming recursion and enable the simple and quick calculation of $\mathcal{C}_{k,l}(\alpha',\alpha)$. These summary statistics are defined relative to the grid points -- so a summary statistic with sub-script $k$ will be based on all the data points for which $x_i\leq g_k$, and we define $n_k$ to be the largest of observation such that $x_{n_k}\leq g_k$.  
\[
S^{(Y)}_k=\sum_{i=1}^{n_k} \frac{y_i}{\sigma_i^2},~~S^{(YY)}_k=\sum_{i=1}^{n_k} \frac{y^2_i}{\sigma_i^2},~~S^{}_k=\sum_{i=1}^{n_k} \frac{1}{\sigma_i^2}
\]
\[
S^{(X)}_k=\sum_{i=1}^{n_k} \frac{x_i}{\sigma_i^2},~~S^{(XX)}_k=\sum_{i=1}^{n_k} \frac{x^2_i}{\sigma_i^2},~~S^{(XY)}_k=\sum_{i=1}^{n_k} \frac{x_iy_i}{\sigma_i^2}.
\]
All summary statistics with sub-script 0, or that involve an empty sum, such that $n_k=0$, are defined to be 0.

If we then define the co-coefficients of $\mathcal{C}_{k,l}(\alpha',\alpha)$, so that
\[
\mathcal{C}_{k,l}(\alpha',\alpha)=A\alpha^2+B\alpha\alpha'+C\alpha+D+E\alpha'+F\alpha'^2,
\]
then tedious algebra gives that these coefficients are defined in terms of the summary statistics as
\begin{eqnarray*}
A&=&\frac{S^{(XX)}_k-S^{(XX)}_l}{(g_k-g_l)^2}-2g_l\frac{S^{(X)}_k-S^{(X)}_l}{(g_k-g_l)^2}+g_l^2\frac{S_k-S_l}{(g_k-g_l)^2},\\
B&=&2(g_k+g_l)\frac{S^{(X)}_k-S^{(X)}_l}{(g_k-g_l)^2}-2\frac{S^{(XX)}_k-S^{(XX)}_l}{(g_k-g_l)^2}-2g_kg_l\frac{S_k-S_l}{(g_k-g_l)^2},\\
C&=&2g_l\frac{S^{(Y)}_k-S^{(Y)}_l}{g_k-g_l}-2\frac{S^{(XY)}_k-S^{(XY)}_l}{g_k-g_l} ,\\ 
D&=&S^{(YY)}_k-S^{(YY)}_l, \\
E&=& 2\frac{S^{(XY)}_k-S^{(XY)}_l}{g_k-g_l}-2g_k\frac{S^{(Y)}_k-S^{(Y)}_l}{g_k-g_l},\\
F&=&\frac{S^{(XX)}_k-S^{(XX)}_l}{(g_k-g_l)^2}-2g_k\frac{S^{(X)}_k-S^{(X)}_l}{(g_k-g_l)^2}+g_k^2\frac{S_k-S_l}{(g_k-g_l)^2}. 
\end{eqnarray*}
Adding $q_{i}^{(k)}(\alpha')+\beta$ to $\mathcal{C}_{k,l}(\alpha',\alpha)$ just changes the coefficients of powers of $\alpha'$ -- that is $D$ increases by $a+\beta$, $E$ increases by $b$ and $F$ increases by $c$. Minimising the resulting quadratic with-respect to $\alpha'$ gives a quadratic of the form $a'+b'\alpha+c'\alpha^2$ where
\[
a'=D+a+\beta-\frac{(E+b)^2}{4(F+c)},
\]
\[
b'=C-\frac{(E+b)B}{2(F+c)},
\]
\[
c'=A-\frac{B^2}{4(F+c)}.
\]

\end{document}